\documentclass[runningheads]{llncs}
\usepackage[T1]{fontenc}
\usepackage{graphicx}
\usepackage{amsmath,amssymb,amsfonts}
\usepackage[linesnumbered,ruled,vlined]{algorithm2e}
\usepackage{subcaption}
\usepackage{multirow}
\usepackage{mathtools}
\usepackage{color,soul}
\usepackage{xcolor}
\usepackage{url} 
\usepackage{hyperref}

\newcommand{\Rev}[1]{\textcolor{black}{ #1}} % Revision comments
\newcommand{\RevReviewer}[1]{\textcolor{black}{#1}}%optional comments
\newcommand{\RevPA}[1]{\textcolor{black}{#1}}%optional comments
\newcommand{\RevHD}[1]{\textcolor{black}{#1}}%optional comments
\newcommand{\RevHDS}[1]{\textcolor{black}{#1}}%optional comments

\definecolor{brightpink}{rgb}{1.0, 0.0, 0.5}

\begin{document}
\title{Direct Exoplanet Detection Using \\ L1 Norm Low-Rank Approximation\thanks{This work was supported by the Fonds de la Recherche Scientifique-FNRS under Grant no T.0001.23. Simon Vary is a beneficiary of the FSR Incoming Post-doctoral Fellowship.}}
%
%\titlerunning{Abbreviated paper title}
% If the paper title is too long for the running head, you can set
% an abbreviated paper title here
%
\author{Hazan Daglayan\inst{1}\orcidID{0009-0006-4843-6913} \and
Simon Vary\inst{1}\orcidID{0000-0002-9929-4357} \and
Valentin Leplat\inst{2}\orcidID{0000-0002-3313-1547}
\and
Nicolas Gillis\inst{3}\orcidID{0000-0001-6423-6897}
\and
P.-A. Absil\inst{1}\orcidID{0000-0003-2946-4178}}
\authorrunning{H. Daglayan et al.}
% First names are abbreviated in the running head.
% If there are more than two authors, 'et al.' is used.
%
\institute{ICTEAM Institute, UCLouvain, Louvain-la-Neuve, Belgium  \email{\{hazan.daglayan,simon.vary,pa.absil\}@uclouvain.be}
\and
Center of Artificial Intelligence Technology, Skoltech, Moscow, Russia \email{V.Leplat@skoltech.ru} \and
Dept.\ of Mathematics and Operational Research, University of Mons, Mons, Belgium \\
\email{nicolas.gillis@umons.ac.be}
}
\maketitle              % typeset the header of the contribution
\begin{abstract}
We propose to use low-rank matrix approximation using the component-wise L1-norm for direct imaging of exoplanets. 
Exoplanet detection \RevHDS{by direct imaging} is a challenging task for three main reasons: (1) the host star is several orders of magnitude brighter than exoplanets, (2) the angular distance between exoplanets and star is usually very small, and (3) \RevHDS{the images are affected by the noises called speckles that} are very similar to the exoplanet signal both in shape and intensity.
We first empirically examine the statistical noise assumptions of the \RevHDS{L1 and L2} models, and then we evaluate the performance of the proposed L1 low-rank approximation (L1-LRA) algorithm based on visual comparisons and receiver operating characteristic (ROC) curves. We compare the results of the L1-LRA with the widely used truncated singular value decomposition (SVD) based on the L2 norm in two different annuli, one close to the star and one far away.

%The abstract should briefly summarize the contents of the paper in 150--250 words.

\end{abstract}
\section{Introduction}
\label{sec:introduction}

In the field of exoplanet detection, the vast majority of planets (about 99\%) have been detected by indirect methods. Over the past decade, we have observed the rapid development of high-contrast imaging as a promising technique for the detection of exoplanets. Although very challenging, direct imaging provides two main advantages compared to indirect methods\RevHDS{. F}irst, \RevHDS{it gives} access to the photons of the planets, so we can obtain information about the atmospheric composition of the planets \cite{Bowler_2018}. Second, %it allows for the detection of planets in a shorter period of time compared to other methods%, thus enabling the detection of planets on wider orbits. 
\RevReviewer{direct imaging allows us to capture images of planets in a matter of hours, while indirect methods can take years, as they often depend on observing \RevHDS{the} orbital motion \RevHDS{of planet} around its host star. This temporal efficiency becomes especially critical when searching for planets with wider orbits. Indirect methods may require an extended duration to detect planets in these regions, while direct imaging opens up opportunities for their discovery.}

%There are three main reasons why very few exoplanets have been observed \RevHDS{by direct imaging}: (1) the small angular distance to the host star due to the distance between the planet and the earth, (2) the high contrast difference between the planet and its host star, and (3) the similarity between noises called \emph{speckles} and planets. 
Due to the small angular separation between the host star and potential exoplanets, direct imaging requires high-resolution images and, therefore, large and high resolution ground-based telescopes. This causes the light to be diffracted by atmospheric turbulence as it passes through the atmosphere. Despite the use of coronagraphs, such as the well-known Lyot coronagraph or, more recently, vortex coronagraphs \cite{Mawet2005}, to reduce the high contrast caused by the brightness of the star and adaptive optics techniques to avoid the aberrations caused by the refraction of the light, \RevHDS{standard signal processing methods may still be unable} to detect the planet in the images, \RevHDS{due to the} residual aberrations \RevHDS{in the form of} quasi-static speckles that are often brighter than the planet and resemble the planet in shape. 

Angular differential imaging (ADI) is a widely used technique in astronomy to reduce the effects of speckle noise in the images \cite{Marois2006Angular}. This technique is based on observations made in pupil tracking mode, in which the star is fixed in the center of the images as the Earth rotates in a night, which causes the exoplanet to rotate around the host with time. ADI aims at building a reference point spread function (PSF) that reproduces a model of the speckle field to be subtracted from the images and aligned  (in some methods, also combined) with the signal of potential exoplanets. 

Several algorithms are used to build the reference PSF in ADI. 
The most popular ones are based on low-rank approximations to build the reference PSF: 
principal component analysis (PCA) \cite{Soummer2012Detection,Amara2012pynpoint}, 
its annular version (AnnPCA) \cite{GomezGonzalez2017VIP,VIP_HCI}, 
non-negative matrix factorization \cite{ren2018non}, 
the local low-rank plus sparse plus Gaussian decomposition (LLSG) \cite{GomezGonzalez2016Lowranka}, 
and more recently low-rank plus sparse trajectory \cite{vary2023low}. Regime-switching model \cite{dahlqvist2020regime} also combines the advantages of numerous PSF subtraction techniques using low-rank approximations. 

Methods based on the low-rank assumption are obtained by transforming the data cube into a matrix such that each frame corresponds to a row of the matrix. In the cases where we fit the low-rank approximation by minimizing the Frobenius norm, \RevHDS{i.e., by PCA (or equivalently, truncated SVD)}, this corresponds to the maximum likelihood estimator under the i.i.d.\ white Gaussian noise assumption. %\ngc{you do not need to define acronyms, here MLE, if you do not use them. In fact, you redefine it later.}
However, several recent lines of work observed that the residual datacube, which is obtained by subtracting the low-rank part from the original data, is more compatible with the Laplacian distribution that has heavier tails instead of Gaussian \cite{Pairet2019STIM,Daglayan2022Likelihood}. 

\RevHDS{In this paper, we propose to perform the low-rank background approximation using the component-wise L1 norm and to apply a data-dependent approximation. This approach leverages two distinct annuli in the frames, one close and the other far from the star. Firstly, it involves an analysis of the data to identify whether the data is better fitted by a Gaussian or Laplacian distribution. Subsequently, we compare the results both visually and through ROC curve analysis.}  

The rest of this paper is structured as follows. In Section \ref{sec:model}, we propose an alternative to PCA in the context of exoplanet detection, namely a component-wise L1 norm low-rank matrix approximation. We investigate different statistical assumptions on the data and analyze the performances, then apply the appropriate low-rank approximation to the data and present the experimental results in Section \ref{sec:num_exp}. Finally, we conclude in Section \ref{sec:conclusion} and 
discuss potential future works. 

\section{\RevHDS{Models and Methods}}%Model Assumptions}
\label{sec:model}

\RevHDS{\subsection{Planet flux estimation}
\label{sec:fluxestimation}}

Let $M\in \mathbb{R}^{T\times N^2 }$ be a matrix of observations that consists of $T$ unfolded frames with size $N \times N$, i.e., 
each row of the matrix represents a single vectorized frame. The model for $M$ proposed in \cite{Daglayan2022Likelihood}\RevHDS{, assuming a single planet located at position $g\in[N]\times[N]$ in the first frame,} is expressed as 
\begin{equation}\label{eq:data_model}
    M = L + a\RevHD{_g}P_g + E, \quad \mathrm{rank}(L)\leq \RevHD{k}, %\quad P_g \in \Lambda, 
\end{equation}
where $L$ is the low-rank background, $E$ is the noise, $a\RevHD{_g}$ is the intensity of the planet referred to as the \emph{flux}, $P_g$ is the planet signature along the trajectory. % from the set of all feasible trajectories $\Lambda$\RevHD{, and
%$g \in [N ]\times [N ]$ denotes the planet's position on the initial frame.}

For such models based on low-rank approximations, the choice for the rank value is crucial.
Indeed, if the rank is too small, the signal of the speckles will remain in the residual \RevHDS{$M-L$}, making it difficult to separate the signal of the planet from the speckles. Conversely, if it is too large, the signal of the planet will be captured by the low-rank matrix, and it will be challenging to find the signal of the planets in the residual.

%% When E is Gaussian
When the error $E$ is Gaussian distributed, the maximum likelihood estimator for $L$ is obtained by the minimization of the Frobenius norm. 
Classical methods, such as AnnPCA and LLSG, fit the low-rank component as follows: 
\begin{align}
    \hat{L} = \underset{L}{\arg\min} \|M - L \|_F \quad 
    \text{\RevHDS{subject to}} \quad \mathrm{rank}(L) \leq \RevHD{k}, 
    \label{pro:PCA}
\end{align}
where $\|A\|_F$ denotes the entry-wise L2-norm of $A$ (the Frobenius norm), which can be solved using the truncated SVD. 
This step is followed by subtracting the low-rank component and identifying the planet $a\RevHD{_g}P_g$ in the residual matrix by 
solving the minimization problem 
\begin{align}
    \hat{a}_g = \arg\min_{a_g > 0}  \left\| M - \hat{L} -a_gP_g\right\|_{2}, \label{pro:l2_planet}
\end{align}
for all possible $P_g \in \Lambda$. 

\RevHDS{In order to detect the presence of a planet,~\eqref{pro:l2_planet} has to be solved for all planet positions $g$ of potential interest, which are all the pixels except for the ones that are too close to the star and those located at the corners since they cannot adequately capture the rotational motion of the planet; see Section 2.2 %~\ref{sec:detection} 
for the detection step. }%However, we exclude pixels corresponding to the star and those located at the corners since they cannot adequately capture the rotational motion of the planet.}

%% Recently E is Laplacian
Recently, it has been observed that the error term \RevHDS{$E$} has heavy tails and more closely follows the Laplacian distribution \cite{Pairet2019STIM}. Consequently, \RevHDS{it was proposed in~}\cite{Daglayan2022Likelihood} to identify the planet using L1 minimization
\begin{align}
    \hat{a}_g = \arg\min_{a_g > 0}  \left\| M - \hat{L} -a_gP_g\right\|_{1}, \label{pro:l1_planet}
\end{align} 
where $\|\cdot\|_1$ is the component-wise L1 norm, that is, $\| M \|_1 = \sum_{i,j} |M[i,j]|$.  

% Noise assumption in IPAS paper inconsistent
%However, this makes the noise assumption in the low-rank speckle subtraction and in the planet identification inconsistent. 
\RevReviewer{However, this approach introduces an inconsistency in the noise assumption between the speckle subtraction \RevHDS{\eqref{pro:PCA}} and planet identification \RevHDS{\eqref{pro:l1_planet}}. While L1 minimization proves effective for planet identification, it raises the question of why not apply the L1 norm to background subtraction as well. %This inconsistency stems from the difference in noise assumptions between the two steps.
}

%\RevReviewer{If L1 norm works well for planet identification, it implies that the noise characteristics in the data align with the assumptions made by the L1 minimization approach. Therefore, it makes sense to explore the application of L1 norm to the background subtraction phase to ensure consistency in noise modeling.}

\RevReviewer{\RevHDS{Consequently}, we propose to fit the low-rank component using the component-wise L1 norm}
%Instead, we propose to fit the low-rank component using the component-wise L1 norm 
\begin{align}
    \hat{L} = \underset{L}{\arg\min} \|M - L \|_1 \quad \mathrm{s.t.}\quad \mathrm{rank}(L) \leq k
    \label{pro:l1}, 
\end{align}
followed by the planet estimation in L1 norm as stated in \eqref{pro:l1_planet}. \RevReviewer{This approach allows us to maintain a consistent noise assumption across both the low-rank speckle subtraction and the planet estimation phases. By doing so, we aim to enhance the overall integrity and reliability of the analysis.}

\RevHDS{In fact, the proposed approach consists of addressing the optimization problem 
\[
\min_{L,a_g} \|M-L-a_gP_g\|_1 \text{ subject to } \mathrm{rank}(L) \leq k
\]
by first minimizing over $L$, then minimizing over $a_g$. (An interesting direction for future work would be to consider alternating minimization. Preliminary results can be found in~\cite{ESANN2023Daglayan}.)}

% L1-LRA is a difficult problem, how to solve it
The L1 low-rank approximation in~\eqref{pro:l1} is an NP-hard problem, even in the rank-one case~\cite{gillis2018complexity}. 
Hence, most algorithms to tackle~\eqref{pro:l1}, such as alternating convex optimization \cite{ke2005alternating}, the Wiberg algorithm \cite{eriksson2010efficient}, and augmented Lagrangian approaches \cite{zheng2012practical}, do not guarantee to find a global optimal solution, unlike in the case of PCA. Moreover, the computed solutions are sensitive to the initialization of the algorithms.  

We use Algorithm~\ref{alg:l1} (L1-LRA) suggested by \cite{gillis2018complexity} to solve \eqref{pro:l1}. It solves the problem using an exact block cyclic coordinate descent method, where the blocks of variables are the columns of $\hat{U}$ and the rows of $\hat{V}$ of the low-rank approximation $\hat{L}=\hat{U}\hat{V}$. 
We initialized the algorithm with the truncated SVD solution, denoted by $\RevHD{\mathrm{H}^\mathrm{SVD}_k({\cdot})}$ in Algorithm~\ref{alg:l1}. 
%\ngc{If you used truncated SVD as the initialization, should't you put in the algorithm?} 
\RevHDS{In our experiments, we apply an annular} version, similar to annular PCA (AnnPCA) \cite{GomezGonzalez2017VIP,VIP_HCI}, \RevHDS{that} selects only the pixels of $M$ in a certain annulus. \RevHDS{Indeed, a}s the \RevHDS{values of pixels} decrease away from the star, it is usually better to calculate the low-rank approximation of each annulus separately.

\begin{algorithm}
    \caption{L1-LRA \cite{gillis2018complexity}\label{alg:l1} }
    \SetAlgoLined
    \KwIn{Image sequence $M \in \mathbb{R}^{t\times n}$, rank $\RevHD{k}$, maximum number of iteration maxiter
    } 

    $\dot{U}, \dot{S}, \dot{V}^T =$ $\RevHD{\mathrm{H}^\mathrm{SVD}_k(M)}$ \\
    $\hat{U} = \dot{U}\dot{S}$; $\quad \hat{V} =\dot{V}^T$\\
    
    \For {$i = $1: \normalfont{maxiter}} {
        $R = M-\hat{U}\hat{V}$ \\
        \For {$\RevHD{j}=1:\RevHD{k}$} {
        $R=R+\hat{U}[:,\RevHD{j}]\hat{V}[\RevHD{j},:]$ \\
        $\hat{U}[:,\RevHD{j}] = \underset{u}{\min}\|R-u \hat{V}[\RevHD{j},:]\|_1$ \\
        $\hat{V}[\RevHD{j},:] =  \underset{v}{\min} \| R^T-v \hat{U}[:,\RevHD{j}]\|_1^T$ \\
        $R=R-\hat{U}[:,\RevHD{j}]\hat{V}[\RevHD{j},:]$ \\
        }
    }
\end{algorithm}
To solve the minimization problem in steps 7-8 of 
Algorithm~\ref{alg:l1}, we use the exact method 
from~\cite{gillis2011dimensionality}; these subproblems are weighted median problems which can be solved in closed form.

\RevHDS{\subsection{Planet detection}}
\label{sec:detection}
%In order to obtain the planet signature, we use likelihood ratio $\Lambda$ map \cite{Daglayan2022Likelihood}, 
\RevHDS{The detection procedure consists of declaring positive the positions $g$ where a detection metric---e.g., an SNR or a likelihood ratio---exceeds a given threshold. A likelihood ratio map $\Lambda$ was proposed in~\cite{Daglayan2022Likelihood},}
which consists of L1 norm likelihood ratios $\Lambda_g(R)$ based on maximizing log-likelihood of Laplace distribution using the solution of \eqref{pro:l2_planet} or \eqref{pro:l1_planet} because it has been shown to provide better results in practice \cite{Daglayan2022Likelihood}
\begin{align}
    \log \Lambda_g(R) = -\!\!\!\!\!\sum_{(t, r) \in {\Omega}_g}\!\!\!\!
		\frac{
			|R(t, r) - \hat{a}_g P_g(t, r)|
			- |{R(t, r)}|
		}{
			\sigma_{R(r)}
		}, 
\label{pro:l1_lr}
\end{align}
where $R = M-\hat{L}$, $\sigma_{R}$ is the standard deviation of $R$ computed along the time dimension, and $\Omega_g$ is the set of indices $(t, r)$ of pixels whose distance from the trajectory is smaller than half the diffraction limit, for more details see \cite{Daglayan2022Likelihood}. 
We will also use the L2 norm $\Lambda$ map which consists of L2 norm likelihood ratios $\Lambda_g(R)$ using the solution of \eqref{pro:l2_planet} or \eqref{pro:l1_planet}
\begin{align}
    \log \Lambda_g(R) = -\frac{1}{2}\sum_{\mathclap{(t, r)\in {\Omega}_g}}^{}
		\frac{
			|R(t, r) - \hat{a}_g P_g(t, r)|^2
			- |{R(t, r)}|^2
		}{
			\sigma^2_{R(r)}
		}
\label{pro:l2_lr}.
\end{align}

\section{Numerical Experiments}
\label{sec:num_exp}

In order to analyze the performance of the L1-LRA algorithm \RevHDS{for background subtraction in exoplanet detection}, we compare the algorithms visually by plotting their log-likelihood detection maps $\Lambda$ and also using receiver operating characteristic (ROC) curves. Additionally, we empirically investigate the methods in terms of fitting the data to Gaussian and Laplacian distributions. The Python codes of the implementations are publicly available from \url{https://github.com/hazandaglayan/l1lra_for_exoplanets}.

We tested the algorithms using the publicly available dataset \emph{sph3} for the exoplanet data challenge \cite{cantalloube2020exoplanet}. The ADI cube obtained with the VLT/SPHERE-IRDIS instrument has 2\RevHD{28} frames with size $160 \times 160$, \RevReviewer{ and it has a total field rotation of
80.5 degrees.} It \RevReviewer{is a real dataset, but it} has 5 synthetically injected planets at different distances from the star.

\subsection{Empirical estimation of the noise distributions}
In order to analyze the suitability of different noise assumptions, we fit the Gaussian and the Laplacian distribution to \RevHDS{the residual data, i.e.,} the data after subtracting the low-rank component using PCA or L1-LRA. We look at two different annuli separately, one that is close to the star at $4\lambda/D$ separation and \RevHDS{one} more distant \RevHDS{from} the star at $10\lambda /D$ and measure the goodness of fit visually and by the \RevPA{\RevHDS{coefficient} of determination $\rho^2$}.

\begin{figure*}[t!]
     \centering
     \begin{subfigure}[b]{0.48\textwidth}
         \centering
         \includegraphics[width=\textwidth]{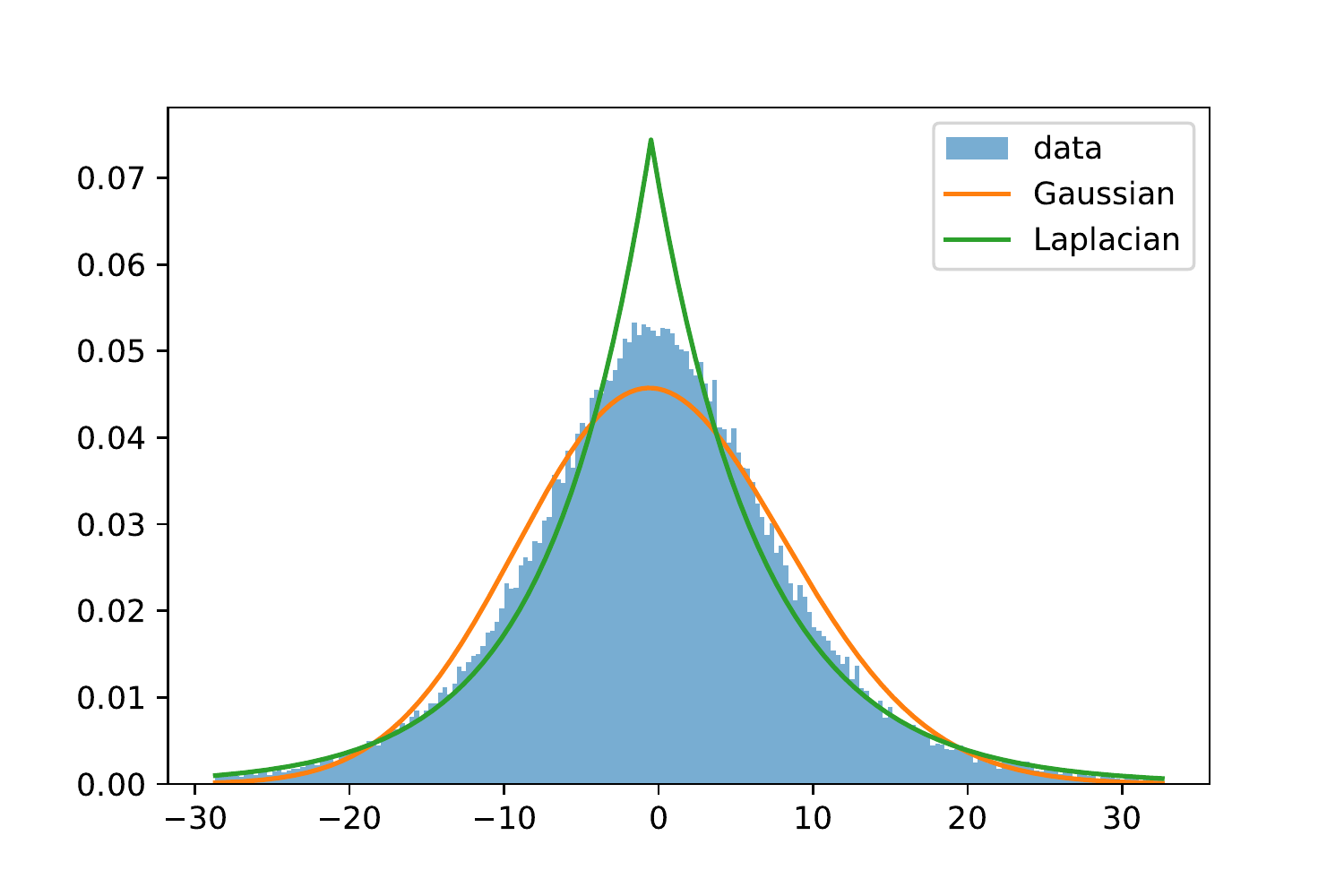}
         \caption{PCA at 4$\lambda/D$ \label{fig:pca_small}}
     \end{subfigure}
    %\qquad
     \begin{subfigure}[b]{0.48\textwidth}
         \centering
         \includegraphics[width=\textwidth]{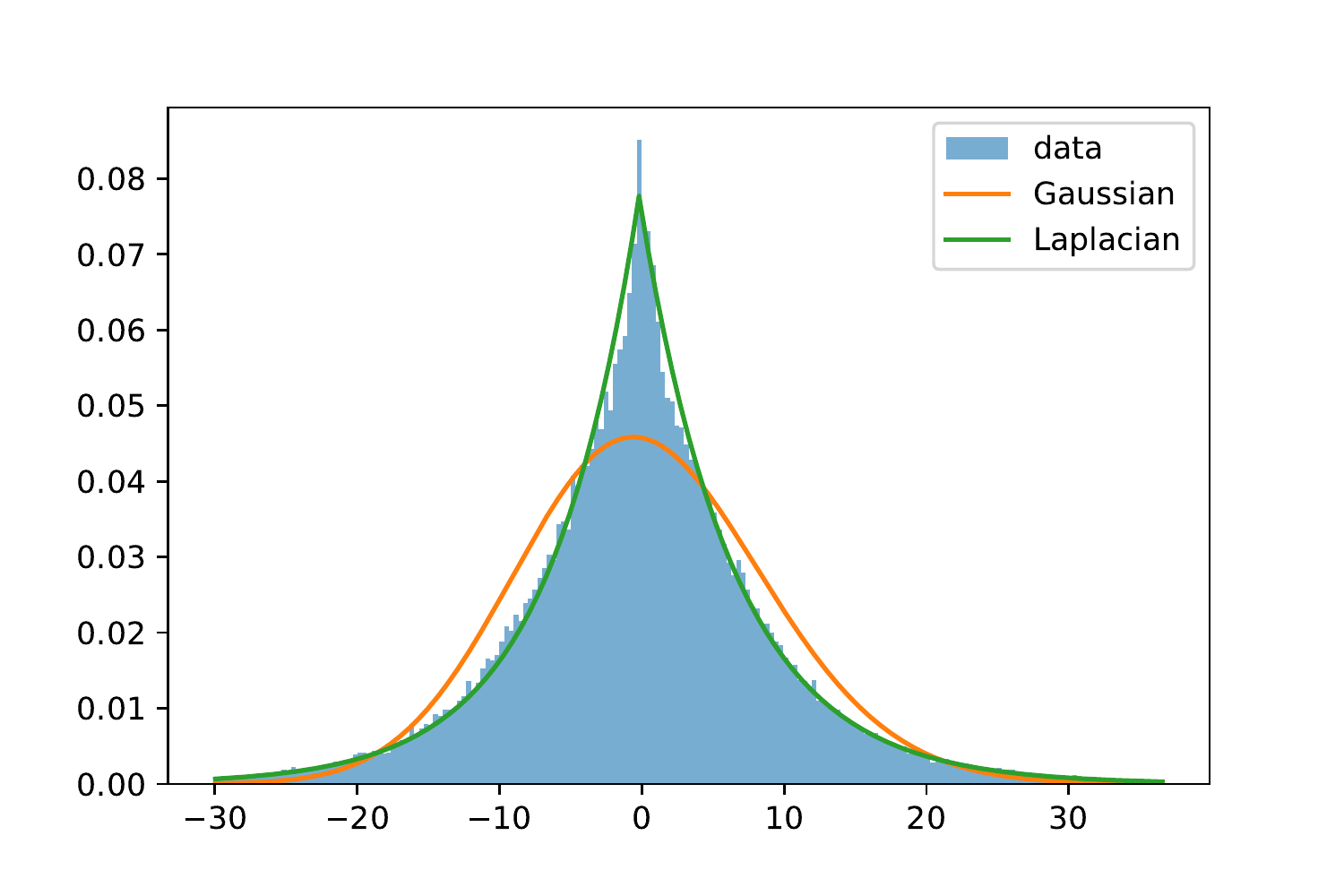}
         \caption{L1-LRA at 4$\lambda/D$ \label{fig:l1_small}}
     \end{subfigure}
     \begin{subfigure}[b]{0.48\textwidth}
         \centering
         \includegraphics[width=\textwidth]{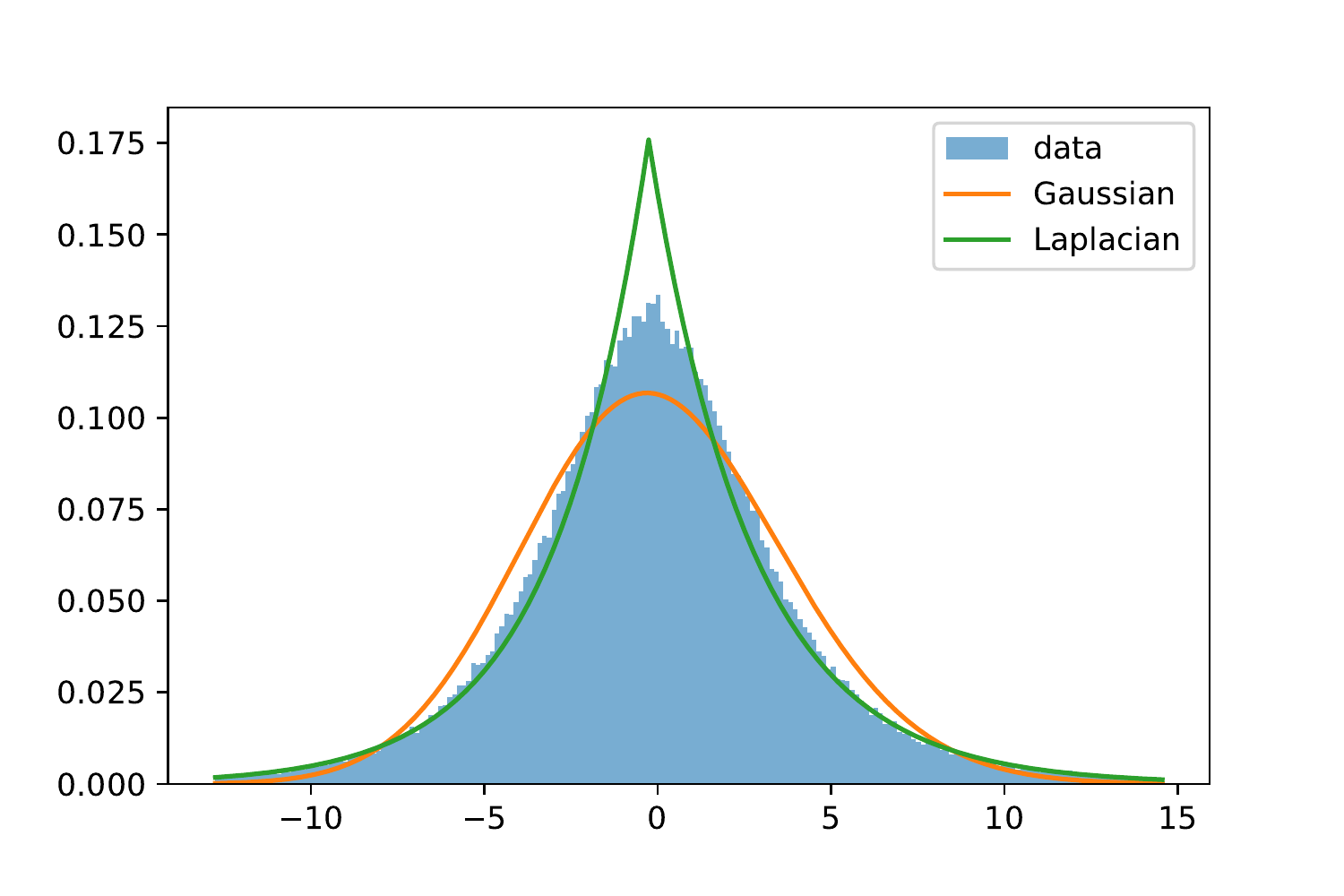}
         \caption{PCA at 10$\lambda/D$ \label{fig:pca_large}}
     \end{subfigure}
    %\qquad
     \begin{subfigure}[b]{0.48\textwidth}
         \centering
         \includegraphics[width=\textwidth]{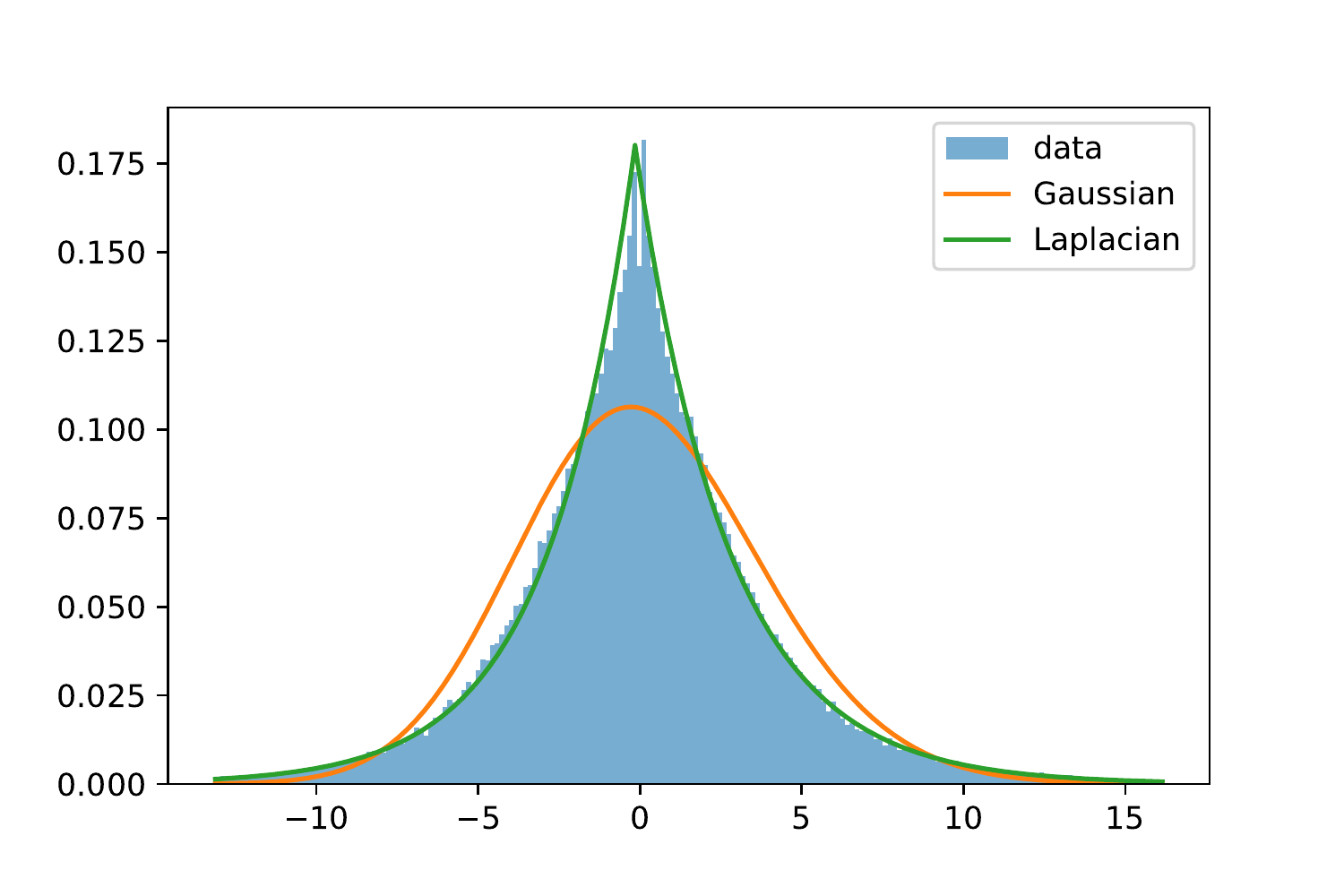}
         \caption{L1-LRA at 10$\lambda/D$ \label{fig:l1_large}}
     \end{subfigure}

        \caption{Residual cube after low-rank approximation applied for small and large separations} \label{fig:distribution}
\end{figure*}

In Figure~\ref{fig:distribution}, we observe \RevPA{that} the residual data follows a  Gaussian distribution after applying PCA and a Laplace distribution after applying L1-LRA in general. \RevPA{However, the Laplacian distribution provides a better fit for the residual cube distribution in the tails, regardless of whether PCA or L1-LRA is used, for both small and large separations.} % Moreover, we see that the distribution after any distribution fits better in the tails for both small and large separations. }
%after applying %any low-rank approximation follows a Laplace distribution in general.%, 
%but we see that the distribution after L1-LRA fits better in the tails for large separations in Fig.~\ref{fig:pca_large}-\ref{fig:l1_large}. Moreover, The distributions in the tails  are quite similar after both PCA and L1-LRA for small separations in Fig.~\ref{fig:pca_small}-\ref{fig:l1_small}.

%Since it is not always easy to see which distribution fits better, we look at the Pearson correlation coefficients ($\rho$) \Rev{of tails} as a metric to measure the performance of the algorithms. 
\RevPA{To assess the quality of fit of a distribution to the residual data, we use the correlation of determination as a metric. The correlation of determination $\rho^2$ quantifies the relationship between two sample sets, denoted as $X=(x_1,\dots,x_n)$ and $Y= (y_1,\dots,y_n)$, and is computed as follows:
\begin{align}
    \rho^2 = \frac{\left(\sum_{i=1}^{n} (x_i -\bar{x})(y_i-\bar{y})\right)^2}{\left(\sum_{i=1}^n(x_i-\bar{x})^2\right)\left(\sum_{i=1}^n(y_i-\bar{y})^2\right)}.
\end{align}
where $\bar{x}$ and $\bar{y}$ represent the means of the variables in samples $X$ and $Y$. This metric aids in comparing the goodness of fit of different distributions. }A \RevPA{$\rho^2$} value close to 1 indicates that the \RevPA{data} is more consistent with the distribution. \RevHDS{In our experiments, $X$ represents the height of the bins of the data histogram, while $Y$ corresponds to the values of the probability density function within these bins.} In Table \RevPA{\ref{tab1:rho_small} and \ref{tab1:rho_large}, we compute $\rho^2$ value between the tails of the data distribution and the Gaussian or Laplacian distributions. T}he highest \RevPA{$\rho^2$} is obtained with \Rev{L1-LRA for both small and} large separations.

\begin{table}[htbp]
\caption{\RevPA{The \RevHDS{coefficient} of determination $\rho^2$} for small separation 4$\lambda/D$ } 
\begin{center}
\begin{tabular}{|c|c|c|c|c|}
\hline
\multirow{2}{*}{\textbf{Rank}} & \multicolumn{2}{c|}{\textbf{PCA}}     & \multicolumn{2}{c|}{\textbf{L1-LRA}}   \\ \cline{2-5} 
                  & \multicolumn{1}{c|}{\textbf{Gaussian}}  & \textbf{Laplacian}  & \multicolumn{1}{l|}{\textbf{Gaussian}}  & \textbf{Laplacian}  \\ \hline
5     & 0.9\RevPA{876} & 0.99\RevPA{18} & 0.9\RevPA{878} & \textbf{0.99\RevPA{20}} \\ \hline
10    & 0.9\RevPA{866} & 0.9\RevPA{890} & 0.9\RevPA{878} & \textbf{0.99\RevPA{38}} \\ \hline
15    & 0.9\RevPA{841} & 0.99\RevPA{48} & 0.99\RevPA{16} & \textbf{0.99\RevPA{72}} \\ \hline
\end{tabular}
\label{tab1:rho_small}
\end{center}
\end{table}

\begin{table}[htbp]
\caption{\RevPA{The \RevHDS{coefficient} of determination $\rho^2$} for large separation 10$\lambda/D$  }
\begin{center}
\begin{tabular}{|c|c|c|c|c|}
\hline
\multirow{2}{*}{\textbf{Rank}} & \multicolumn{2}{c|}{\textbf{PCA}}     & \multicolumn{2}{c|}{\textbf{L1-LRA}}   \\ \cline{2-5} 
                  & \multicolumn{1}{c|}{\textbf{Gaussian}}  & \textbf{Laplacian}  & \multicolumn{1}{l|}{\textbf{Gaussian}}  & \textbf{Laplacian}  \\ \hline
5     & 0.9\RevPA{866} & 0.99\RevPA{20} & 0.9\RevPA{894} & \textbf{0.99\RevPA{40}} \\ \hline
10    & 0.9\RevPA{859} & 0.99\RevPA{36} & 0.9\RevPA{872} & \textbf{0.99\RevPA{48}} \\ \hline
15    & 0.99\RevPA{12} & 0.99\RevPA{54} & 0.99\RevPA{22} & \textbf{0.99\RevPA{6}0} \\ \hline
\end{tabular}
\label{tab1:rho_large}
\end{center}
\end{table}

\begin{figure*}[t!]
     \centering
          \begin{subfigure}[b]{1\textwidth}
            \centering
            \begin{subfigure}[b]{0.25\textwidth}\includegraphics[width=\textwidth]{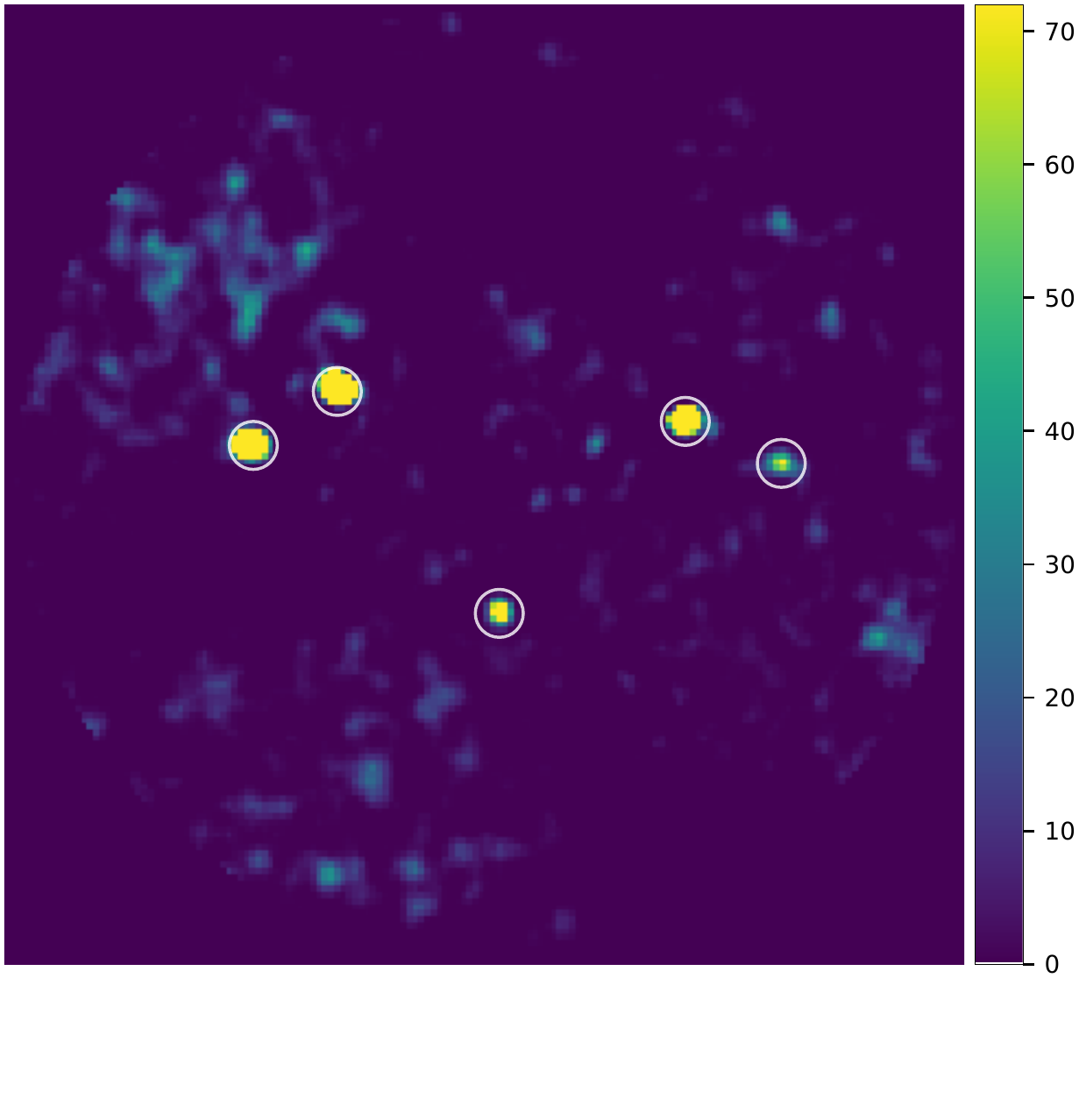}
            \end{subfigure}
            \begin{subfigure}[b]{0.45\textwidth}\includegraphics[width=\textwidth]{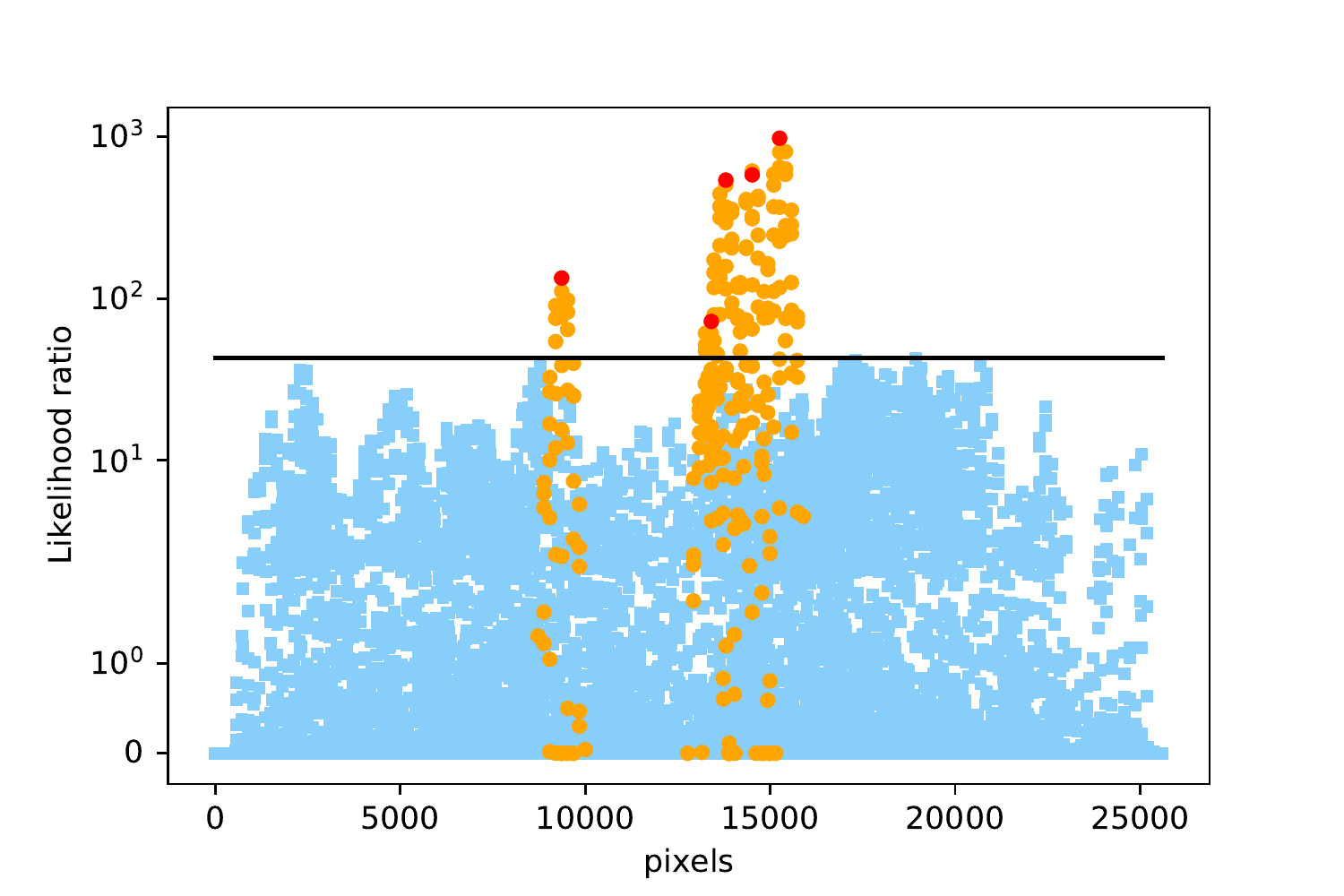}
            \end{subfigure}
            
            %\caption{L1L1\label{fig:irdis_l1}}
     \end{subfigure}
     \begin{subfigure}[b]{1\textwidth}
            \centering
            \begin{subfigure}[b]{0.25\textwidth}\includegraphics[width=\textwidth]{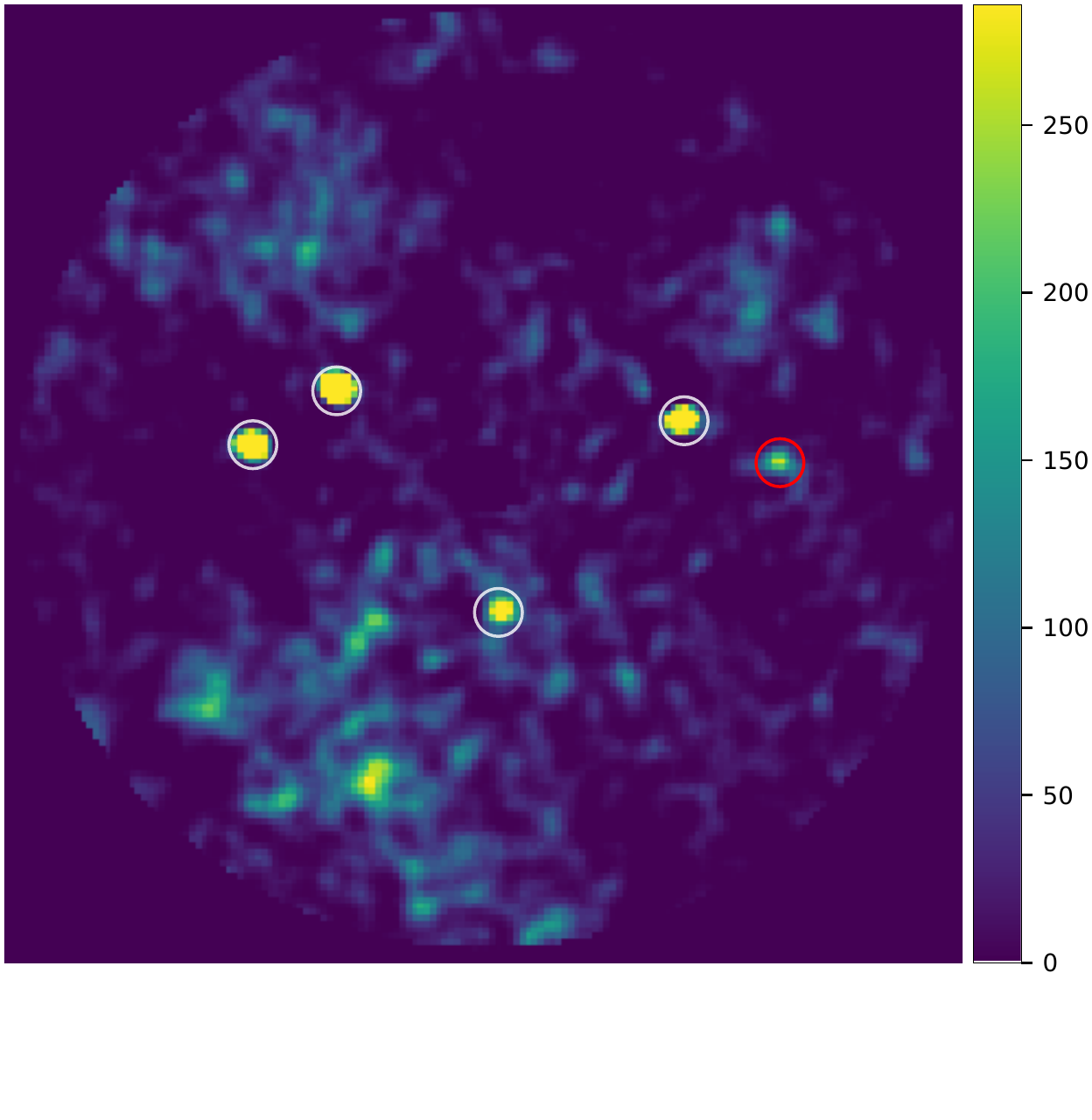}
            \end{subfigure}
            \begin{subfigure}[b]{0.45\textwidth}\includegraphics[width=\textwidth]{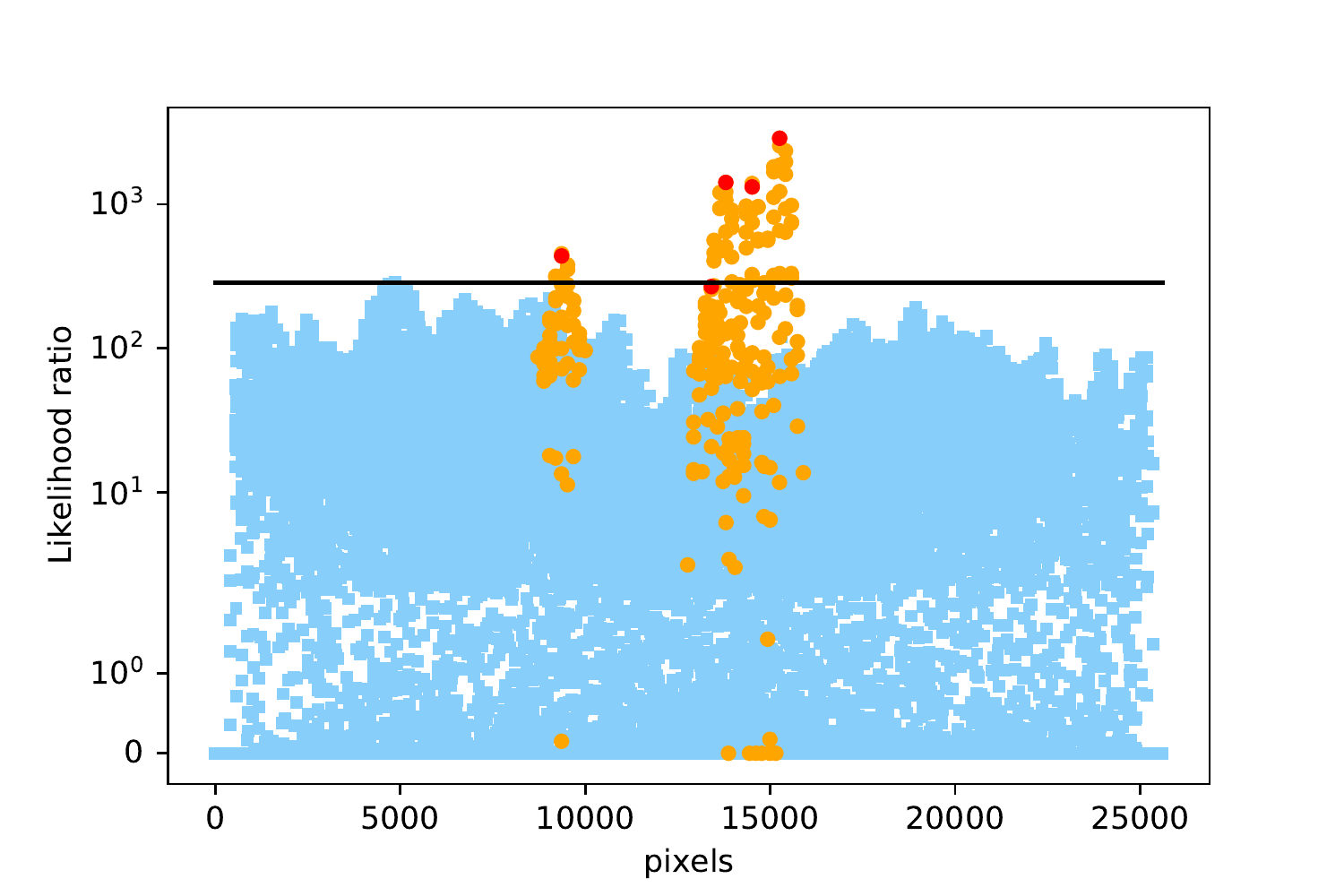}
            \end{subfigure}
            %\caption{L1L2\label{fig:irdis_l1l2}}
     \end{subfigure}
     \begin{subfigure}[b]{1\textwidth}
            \centering
            \begin{subfigure}[b]{0.25\textwidth}\includegraphics[width=\textwidth]{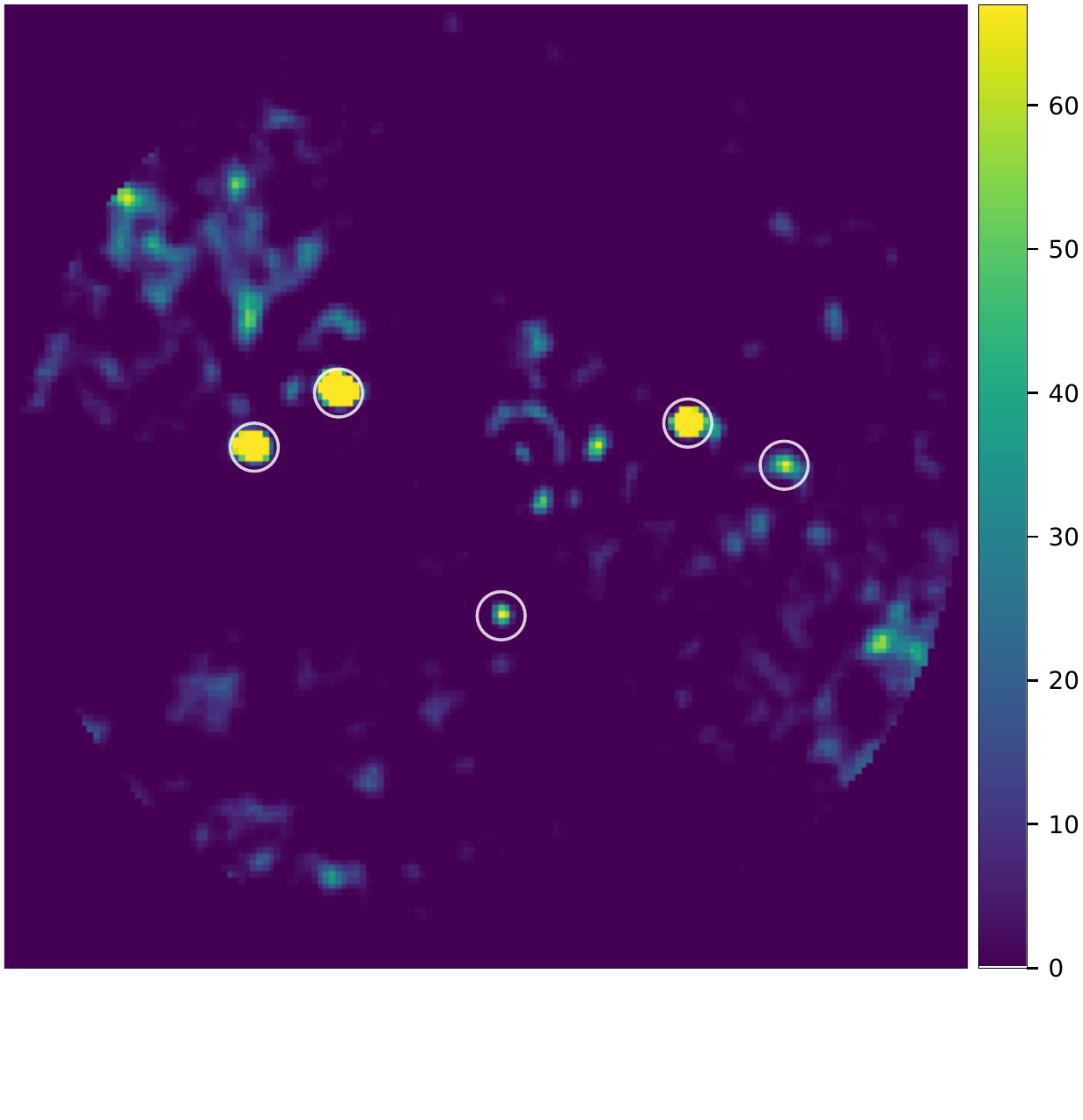}
            \end{subfigure}
            \begin{subfigure}[b]{0.45\textwidth}\includegraphics[width=\textwidth]{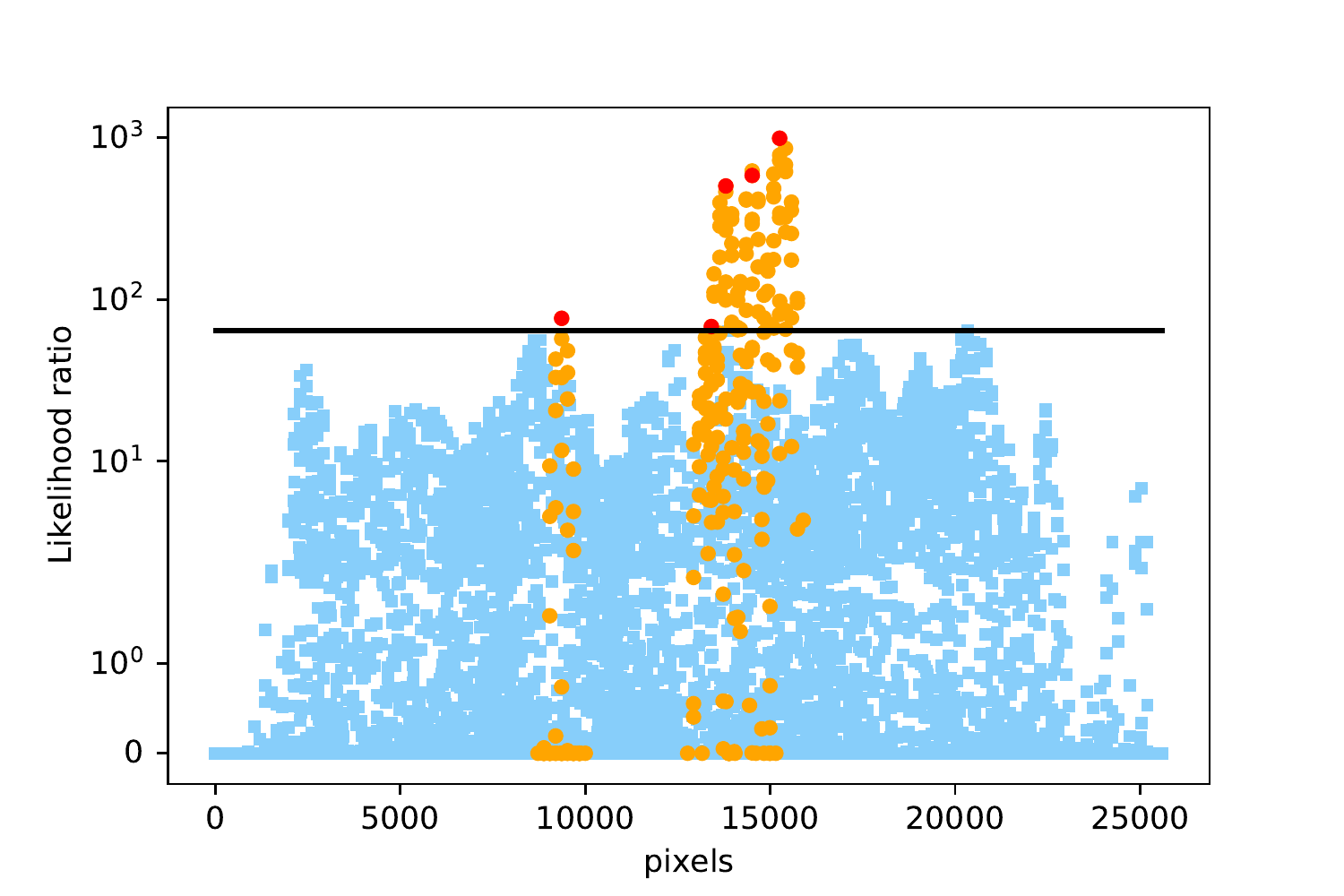}
            \end{subfigure}
            %\caption{L2L1\label{fig:irdis_l21l1}}
     \end{subfigure}
     \begin{subfigure}[b]{1\textwidth}
            \centering
            \begin{subfigure}[b]{0.25\textwidth}\includegraphics[width=\textwidth]{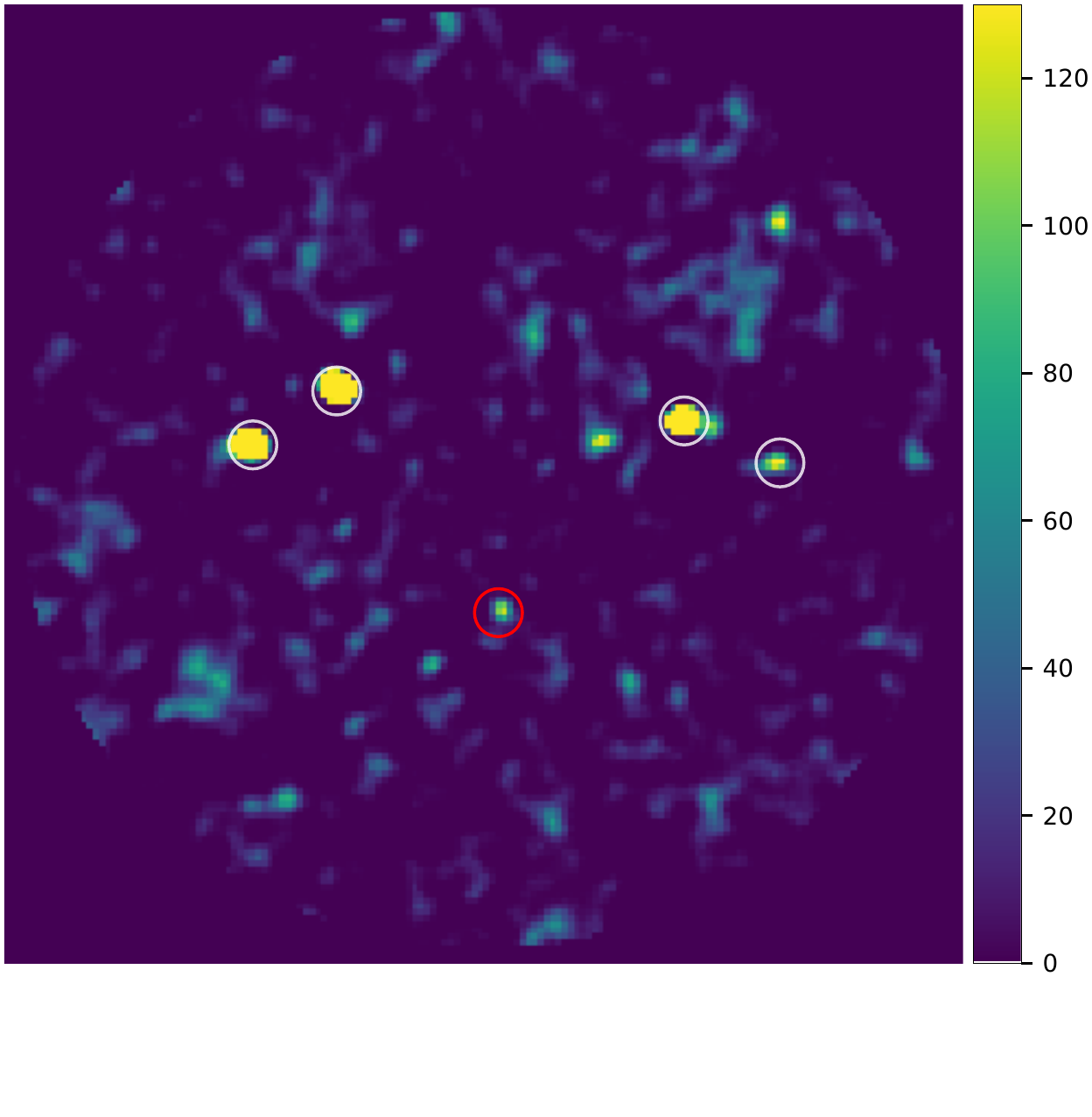}
            \end{subfigure}
            \begin{subfigure}[b]{0.45\textwidth}\includegraphics[width=\textwidth]{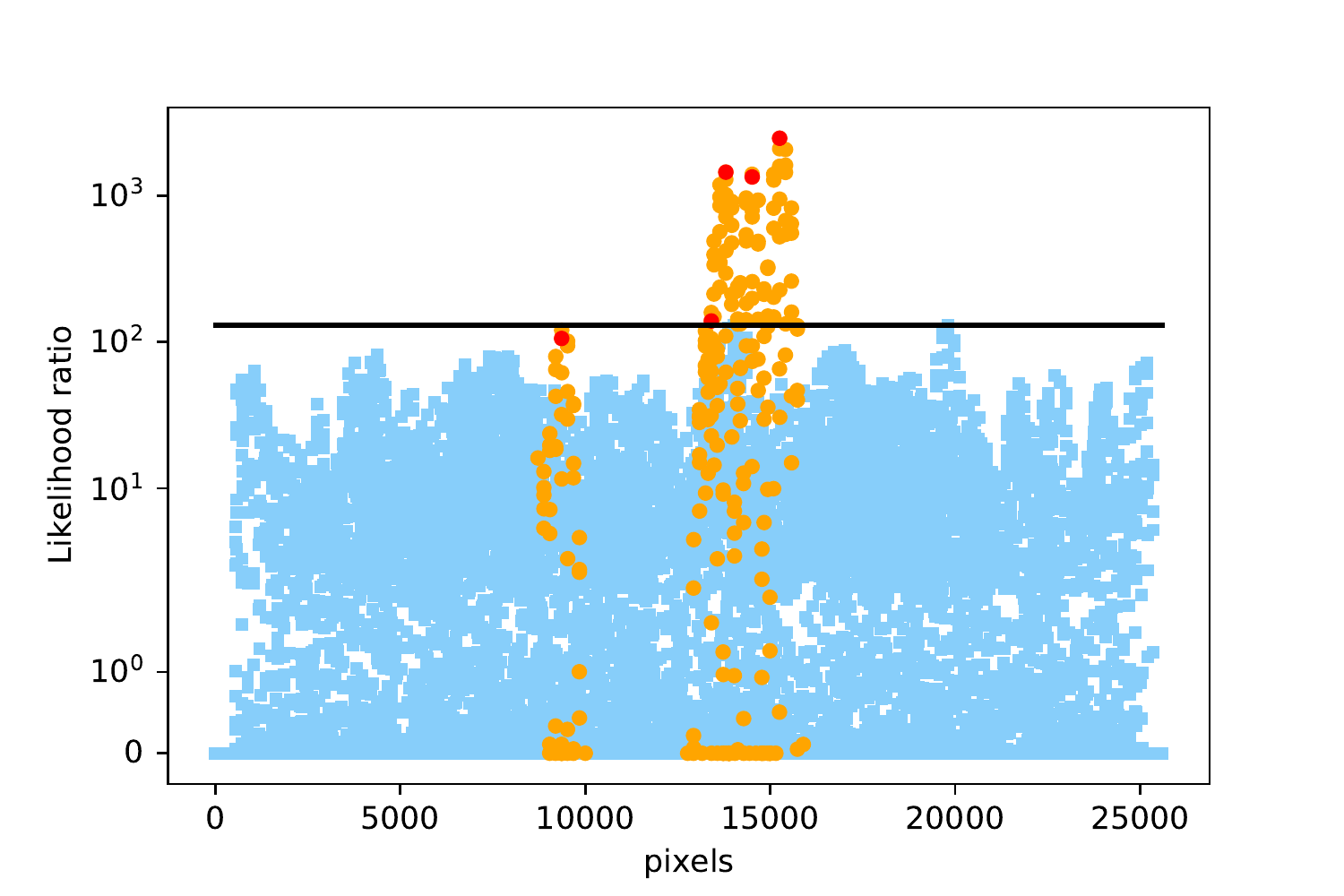}
            \end{subfigure}
            %\caption{L2L2\label{fig:irdis_l2}}
     \end{subfigure}
        \caption{Left column: $\Lambda$ map results of the methods L1L1, L1L2, L2L1, and L2L2 from top to bottom. The \RevHDS{detected planets} %locations of the injected planets 
        are circled in white \RevHDS{and the non-detected planets are circled in red}. Right column: Intensity of the pixels in the $\Lambda$ maps. Those maps have been flattened such that the horizontal axis corresponds to the pixels of the $\Lambda$ map, and the vertical axis indicates the intensity of the pixels in the $\Lambda$ map. The center pixels of the planets and the pixels near the planet are shown in thick red and orange dots, respectively. Other pixels are displayed as blue squares. The black line corresponds to the highest blue square. }\label{fig:sphere_detection}
\end{figure*}

\subsection{Performance comparison of L1-LRA and PCA}
We test the performance advantage of using the L1-LRA instead of PCA approximation in terms of visual quality and ROC curves tested on the sph3 test data~\cite{cantalloube2020exoplanet}. We also provide ablation studies using combinations of mixed models, where the low-rank $L$ is first subtracted from the original data $M$ using L1-LRA or PCA, then used in the planet detection. As such, we have four possible algorithms to investigate; see Table~\ref{tab:algorithms}.
\begin{table}[htbp]
\caption{Descriptions of algorithms}
\begin{center}
\begin{tabular}{|c|c|c|}
\hline
\textbf{}& \multicolumn{2}{|c|}{\textbf{Planet detection}} \\
\cline{2-3}
\textbf{Background subtraction}& L1 \textbf{\eqref{pro:l1_planet}} and \textbf{\eqref{pro:l1_lr}} & L2 \textbf{\eqref{pro:l2_planet}} and \textbf{\eqref{pro:l2_lr}} \\
\hline
L1-LRA \textbf{\eqref{pro:l1}} & \quad L1L1 \quad & \quad L1L2 \quad \\
\hline
PCA \textbf{\eqref{pro:PCA}} & \quad L2L1 \quad & \quad L2L2 \quad \\
\hline
\end{tabular}
\label{tab:algorithms}
\end{center}
\end{table}

%We define the L1L1 algorithm as first solving \eqref{pro:l1} for background subtraction and then for each trajectory, followed by solving \eqref{pro:l1_planet} and using $\hat{a}_g$ in \eqref{pro:l1_lr} to detect planet, and so on for other algorithms.

We first compared the algorithms, %\RevHDS{L1L1, L1L2, L2L1, and L2L2,} 
visually in the likelihood ratio map $\Lambda$ map using the dataset with five synthetic planets. We selected the best performing rank according to the average \RevHDS{likelihood ratio} over the locations of the injected planets. Based on this, we chose ranks 5, 6, 5, and 10 for L1L1, L1L2, L2L1, and L2L2, respectively.

Figure~\ref{fig:sphere_detection} shows the $\Lambda$ map and the intensity of pixels in the $\Lambda$ map obtained with the four algorithms. In each right-hand plot, the black line shows the lowest threshold for which there is no false positive. At this threshold, we observe that L1L1 and L2L1 detect the five injected planets, whereas L1L2 and L2L2 miss one of them. Moreover, the likelihood ratios of the planet pixels (the red dots in the figures) obtained by L1L1 are much larger than the threshold.%of \RevHDS{the} algorithms. We observe that \RevHDS{L1L1 and L2L1} algorithms managed to identify the correct locations of the \RevHDS{all planets} with the detection counted above a threshold of \RevHDS{43} for \RevHDS{L1L1}, and \RevHDS{64} for L2L1, which are decided according to the \RevHDS{the highest threshold for which there is no false negative} for each algorithm. \RevHDS{Similarly, we decide the threshold 285 and 130 for L1L2 and L2L2. However, they can not detect one planet each when we choose these thresholds.} In general, we see that \RevHDS{L1L2 and } L2L2 is much more prone to false positives, while L1L1 \RevHDS{L1L2} ha\RevHDS{ve} no false detections because there are no pixels not belonging to an exoplanet (in blue on Figure~\ref{fig:sphere_detection}) above the threshold. \RevHDS{Moreover, the likelihood ratios on the planet pixels, i.e., red dots in the figures, obtained by L1L1 are much larger than the threshold.}

\begin{figure*}[t!]
     \centering
     \begin{subfigure}[b]{0.49\textwidth}
         \centering
         \includegraphics[width=\textwidth]{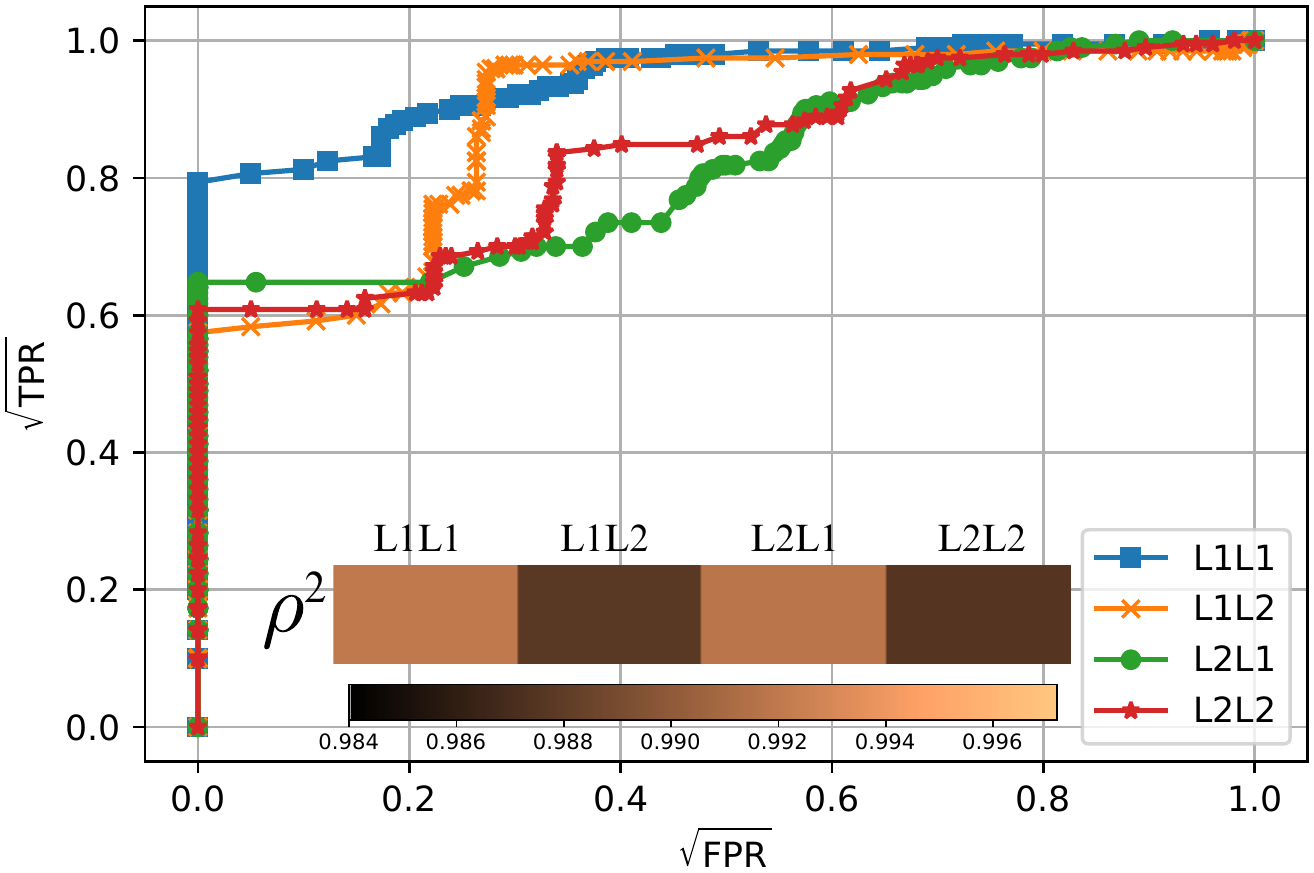}
         %\caption{}
     \end{subfigure}
     \begin{subfigure}[b]{0.49\textwidth}
         \centering
         \includegraphics[width=\textwidth]{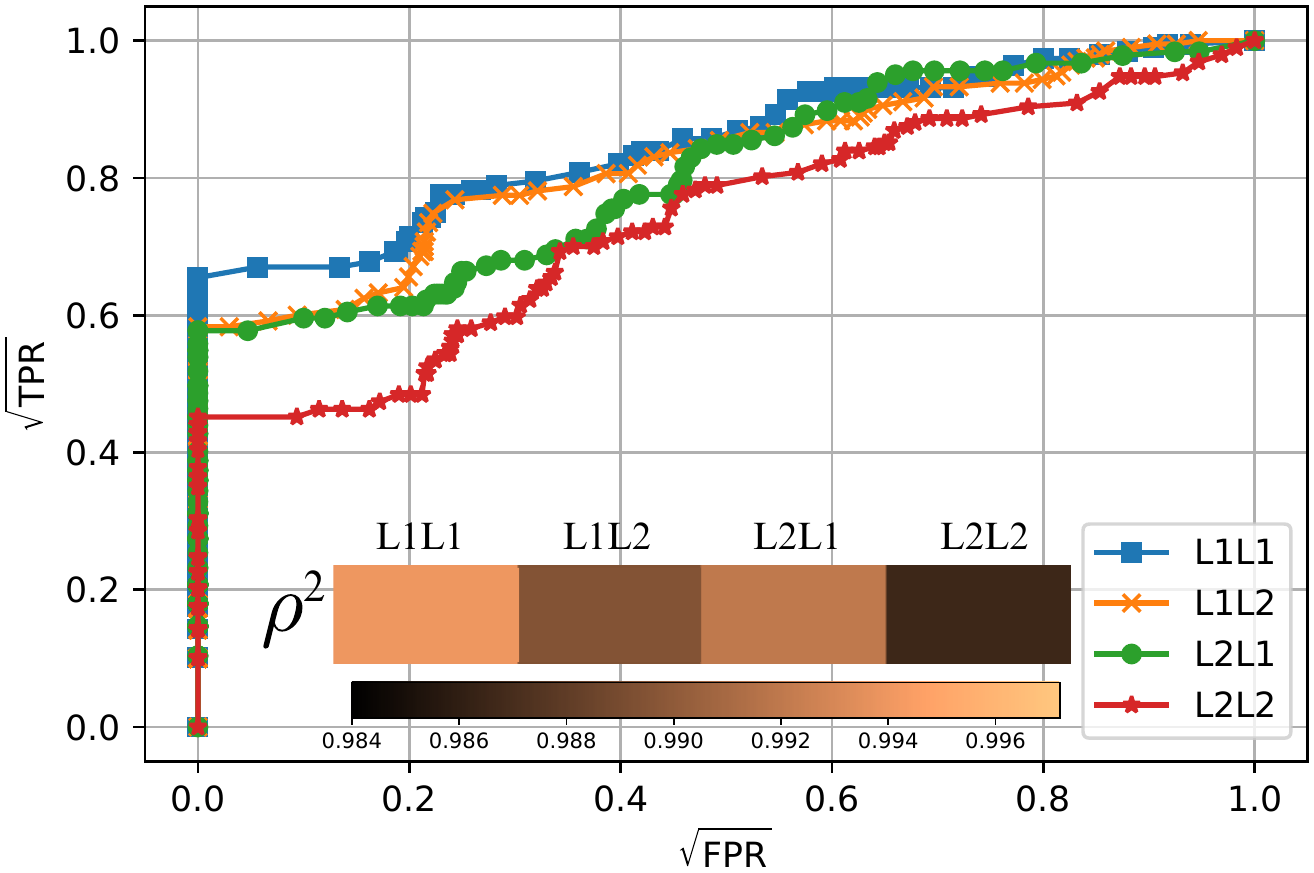}
         %\caption{}
     \end{subfigure}
     \begin{subfigure}[b]{0.49\textwidth}
         \centering
         \includegraphics[width=\textwidth]{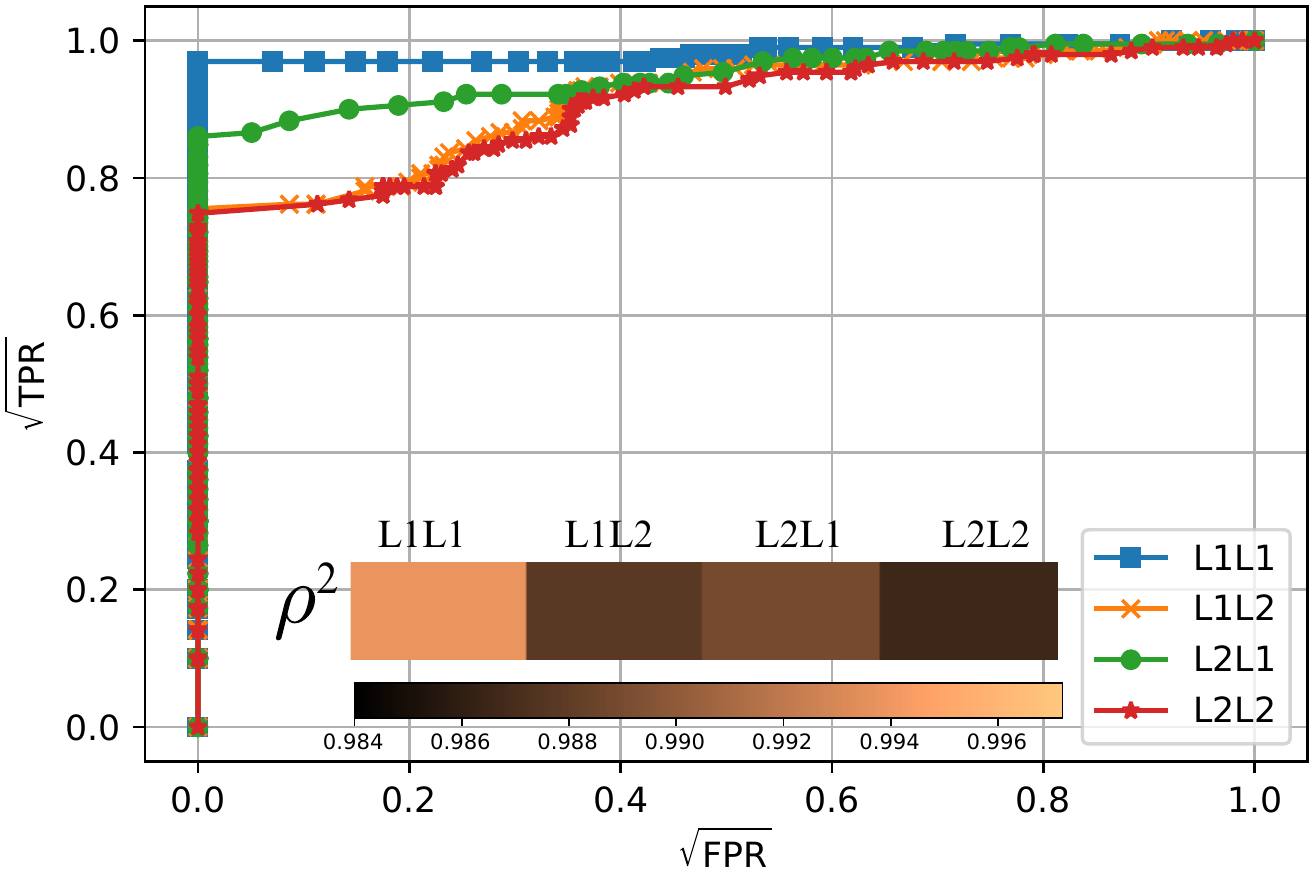}
         %\caption{FROC curve\label{fig:froc}}
     \end{subfigure}
     \begin{subfigure}[b]{0.49\textwidth}
         \centering
         \includegraphics[width=\textwidth]{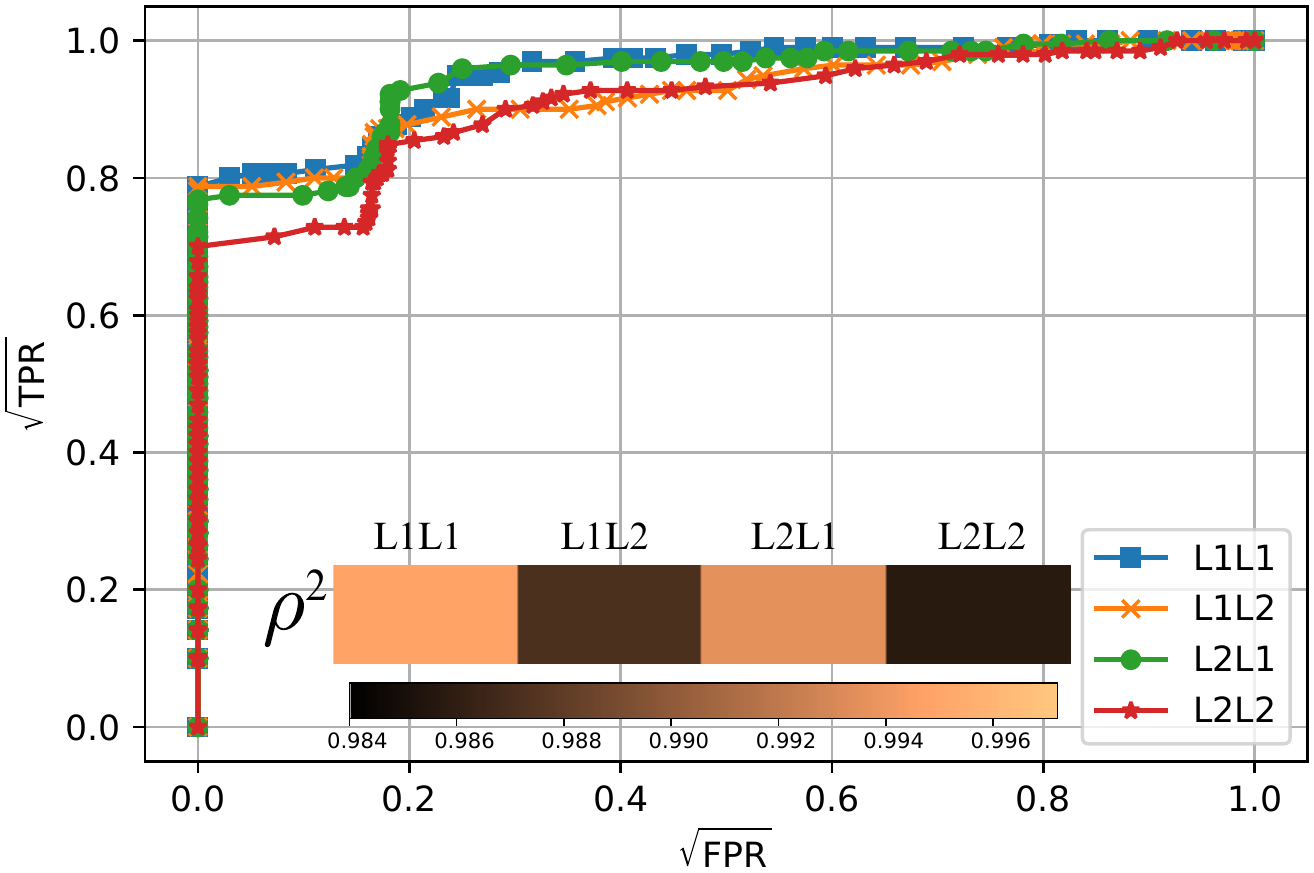}
         %\caption{FROC curve\label{fig:froc}}
     \end{subfigure}
    \begin{subfigure}[b]{0.49\textwidth}
         \centering
         \includegraphics[width=\textwidth]{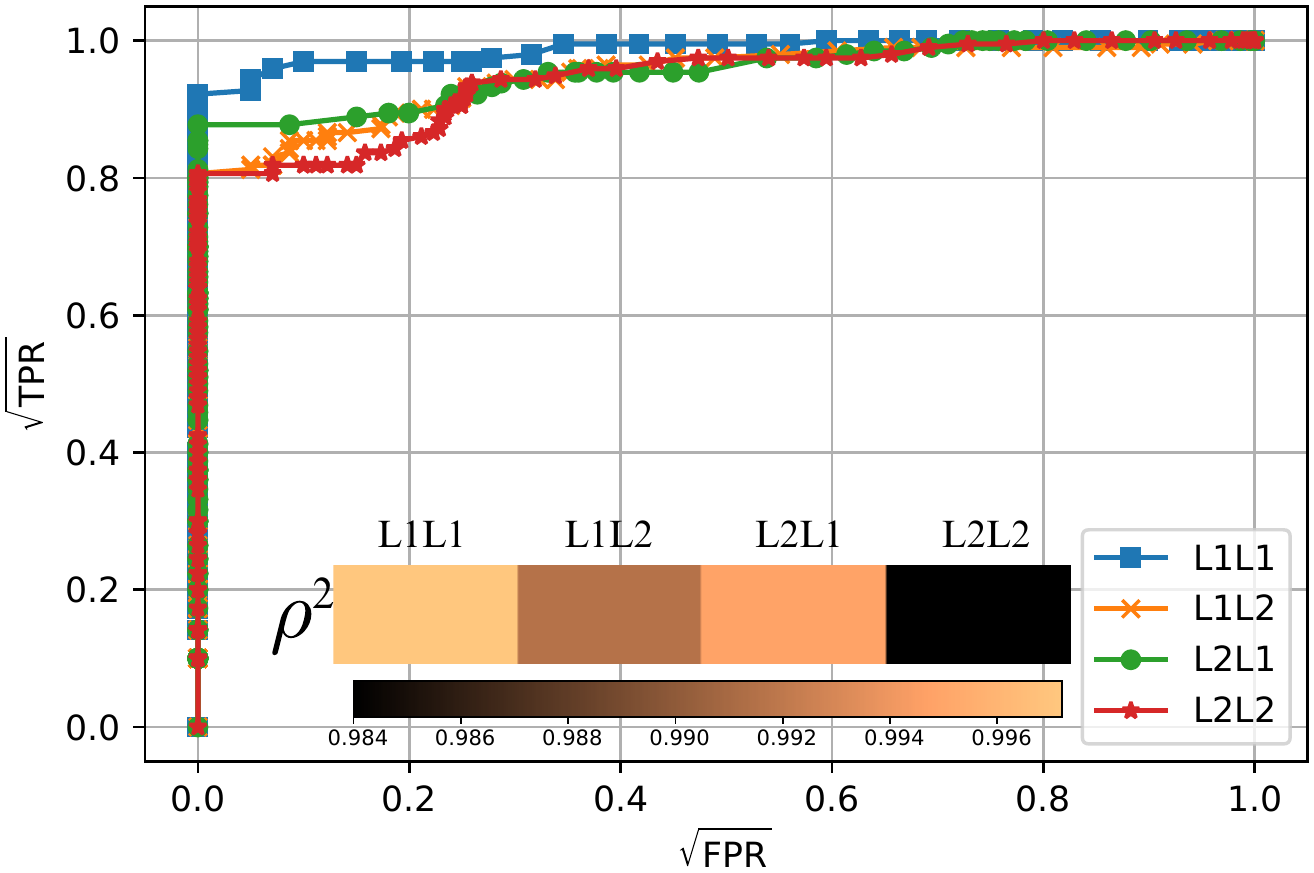}
         %\caption{FROC curve\label{fig:froc}}
     \end{subfigure}
    \begin{subfigure}[b]{0.49\textwidth}
         \centering
         \includegraphics[width=\textwidth]{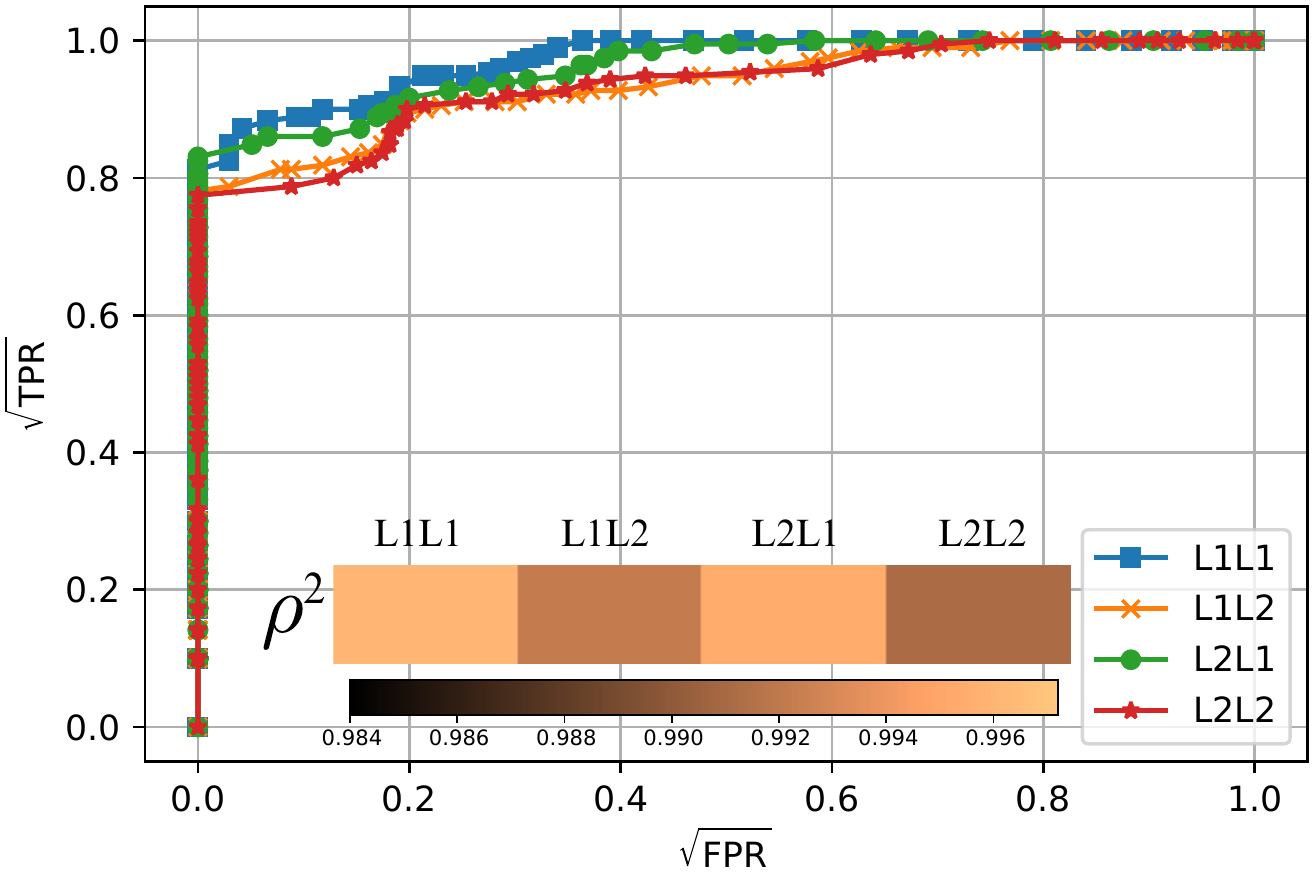}
         %\caption{FROC curve\label{fig:froc}}
     \end{subfigure}
        \caption{\Rev{ROC curves. Left column plots belong to data cubes where the planets are injected in small separation (4$\lambda/D$). Right column plots belong to data cubes where the planets are injected in large separation (10$\lambda/D$).}\RevHDS{The rank is 5, 10, and 15 from top to bottom.  Brown bars show the coefficient of determination $\rho^2$ for each algorithm. }}\label{fig:rocs}
\end{figure*}

To evaluate the performance of the four methods, we put them to the test using synthetically generated data and examine their results using ROC curves in Fig. \ref{fig:rocs}. The detection map used as input for the ROC curve procedure of \cite{Daglayan2022Likelihood} is the $\Lambda$ map. We deleted the five injected planets from the dataset using VIP-HCI package \cite{GomezGonzalez2017VIP,VIP_HCI}. Then, we created \Rev{50} different datasets by injecting \Rev{two} planets in each, \Rev{180} degrees apart, and placed at the separation 4$\lambda/D$. We set the intensity of each injected planet as \Rev{$1\sigma$} where $\sigma$ is the standard deviation of the annulus. We applied the same procedure to a larger separation of $10\lambda/D$ \Rev{with the intensity $0.6\sigma$}. %We compared the results of L1L1, L1L2, L2L1, and L2L2 by using ROC curves. The results are tested for three different rank values $\RevHD{k} = \{5, 10, 15\}$. 
The four methods were tested for three different rank values $k=\{5,10,15\}$. 

In the ROC curve results, we focus on the number of true positives before the first false positives are found, as done in \cite{Daglayan2022Likelihood}.  In the ROC curves obtained by injecting the planets at the small separation (left-hand column of Figure~\ref{fig:rocs}), L1L1 always gives the best results. This is also the case for the large separation with $k=5$ (top right plot of Figure~\ref{fig:rocs}). In the remaining two plots (i.e., large separation and $k=\{10,15\}$), the four algorithms perform rather similarly. %Figure \ref{fig:rocs} shows the ROC curves obtained by injecting \RevHDS{at} a small separation in the first column. In \Rev{all these} ROC curves, \Rev{L1L1} gives the best results. In the ROC curve of the results of $\RevHD{k}=5$ we obtained by injecting into the large separation, L1L1 outperforms all other methods in terms of the ROC curves\Rev{, but in the results of $\RevHD{k}=\{10, 15\}$, the algorithms have similar performances}.
\Rev{ Moreover, we can see a relationship between the ROC curve performances of the algorithms and the $\RevPA{\rho^2}$ values. L1L1, which corresponds to blue curves, has the highest $\RevPA{\rho^2}$ value, while L2L2, which corresponds to red curves, has the lowest value. In addition, L2L1 generally has the second-highest $\RevPA{\rho^2}$ value while generally performing the second-best on ROC curves. }

\begin{figure*}[t!]
     \centering
     \begin{subfigure}[b]{0.49\textwidth}
         \centering
         \includegraphics[width=\textwidth]{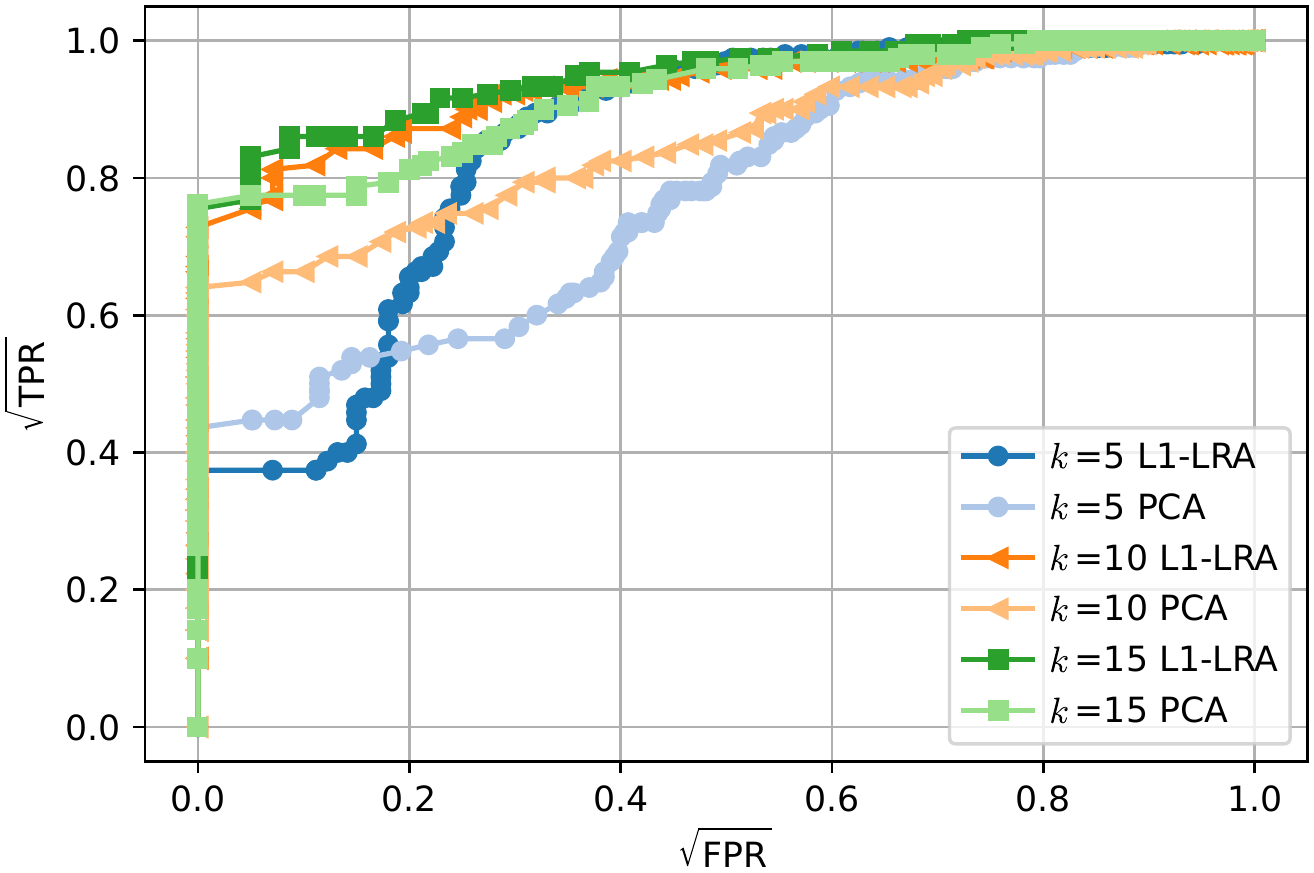}
         %\caption{}
     \end{subfigure}
     \begin{subfigure}[b]{0.49\textwidth}
         \centering
         \includegraphics[width=\textwidth]{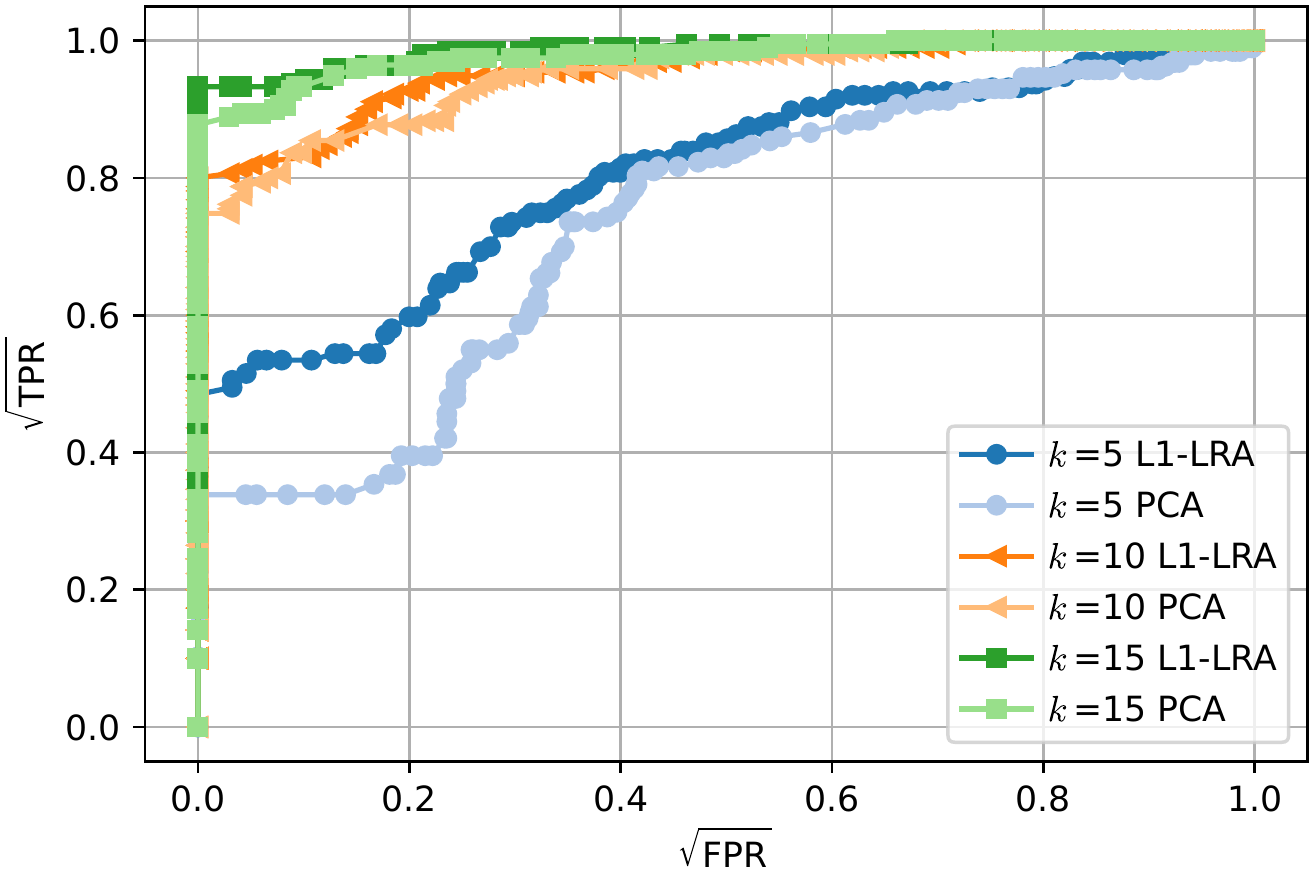}
         %\caption{}
     \end{subfigure}
        \caption{\Rev{ROC curves of SNR maps. The left column plot belongs to data cubes where the planets are injected in small separation (4$\lambda/D$), while the right column plot belongs to data cubes where the planets are injected in large separation (10$\lambda/D$).}}\label{fig:rocs_snr}
\end{figure*}

\Rev{We also compared the L1-LRA and PCA algorithms using SNR map as a detection map, which is a classical approach in direct imaging \cite{Mawet_2014}. We use the implementation of SNR available in the VIP package~\cite{GomezGonzalez2017VIP,VIP_HCI}. Figure \ref{fig:rocs_snr} shows the ROC curves of SNR maps applied after PCA and L1-LRA. We use the same inputs and procedure as described above. Except for the results of $\RevHD{k}=5$, L1-LRA outperforms PCA in all ROC curves for both separations. }

\section{Conclusion}
\label{sec:conclusion}
In this study, we proposed a low-rank approximation algorithm in the sense of the component-wise L1-norm, termed L1-LRA, \RevHDS{for background subtraction to detect exoplanets}. A comparison with the commonly used PCA based on the L2 norm was conducted within two distinct annuli: one near the host star and the other farther away. Through empirical analysis, we examined the statistical noise assumptions of our models and evaluated the performance of the L1-LRA algorithm using visual comparisons and ROC curves. In addition to the success observed in these comparisons, our evaluation of performance metrics, including \RevPA{the coefficient of determination $\rho^2$}, indicates that \RevPA{the} L1L1 approach consistently outperforms alternatives. The proposed L1-LRA algorithm presents a promising solution for the challenging task of exoplanet detection. 

There are several directions for future studies, such as using different initializations for the L1 norm approximation or designing more efficient algorithms (e.g., based on smoothing techniques).
\RevPA{In concurrent work~\cite{ESANN2023Daglayan}, we have investigated a technique that consists in estimating both $L$ and $a_g$ simultaneously in~\eqref{eq:data_model}.}

\bibliographystyle{splncs04}
\bibliography{refs}

%\begin{thebibliography}{8}
%\bibitem{ref_article1}
%Author, F.: Article title. Journal \textbf{2}(5), 99--110 (2016)

%\bibitem{ref_lncs1}
%Author, F., Author, S.: Title of a proceedings paper. In: Editor,
%F., Editor, S. (eds.) CONFERENCE 2016, LNCS, vol. 9999, pp. 1--13.
%Springer, Heidelberg (2016). \doi{10.10007/1234567890}

%\bibitem{ref_book1}
%Author, F., Author, S., Author, T.: Book title. 2nd edn. Publisher, Location (1999)

%\bibitem{ref_proc1}
%Author, A.-B.: Contribution title. In: 9th International Proceedings on Proceedings, pp. 1--2. Publisher, Location (2010)

%\bibitem{ref_url1}
%LNCS Homepage, \url{http://www.springer.com/lncs}. Last accessed 4 Oct 2017
%\end{thebibliography}
\end{document}